\documentclass[aps,pra,reprint,longbibliography,superscriptaddress]{revtex4-1}
\usepackage[dvipdfmx]{graphicx} 
\usepackage{color}
\usepackage{amsmath, amssymb}
\usepackage{longtable}
\usepackage{natbib}
\usepackage{bm}
\usepackage{hyperref}
\usepackage{mwe}
\usepackage{multirow}

\allowdisplaybreaks[1]

\begin{document}
\title{
Ferromagnetic kinetic exchange interaction in magnetic insulators
}
\author{Zhishuo Huang}
\affiliation{Theory of Nanomaterials Group, KU Leuven, Celestijnenlaan 200F, B-3001 Leuven, Belgium}
\author{Dan Liu}
\affiliation{Institute of Flexible Electronics, Northwestern Polytechnical University, 127 West Youyi Road, Xi'an, 710072, Shaanxi, China}
\affiliation{Theory of Nanomaterials Group, KU Leuven, Celestijnenlaan 200F, B-3001 Leuven, Belgium}
\author{Akseli Mansikkam\"{a}ki}
\affiliation{NMR Research Unit, University of Oulu, P.O. Box 3000, FI-90014 Oulu, Finland}
\affiliation{Department of Chemistry, Nanoscience Centre, University of Jyv\"{a}skyl\"{a}, FI-40014 University of Jyv\"{a}skyl\"{a}, Finland}
\author{Veacheslav Vieru}
\affiliation{Maastricht Science Programme, Faculty of Science and Engineering, Maastricht University, Paul-Henri Spaaklaan 1, 6229 EN Maastricht, The Netherlands}
\affiliation{Theory of Nanomaterials Group, KU Leuven, Celestijnenlaan 200F, B-3001 Leuven, Belgium}
\author{Naoya Iwahara}
\email{naoya.iwahara@gmail.com}
\affiliation{Department of Chemistry, National University of Singapore, Block S8 Level 3, 3 Science Drive 3, 117543, Singapore}
\affiliation{Theory of Nanomaterials Group, KU Leuven, Celestijnenlaan 200F, B-3001 Leuven, Belgium}
\author{Liviu F. Chibotaru}
\email{liviu.chibotaru@kuleuven.be}
\affiliation{Theory of Nanomaterials Group, KU Leuven, Celestijnenlaan 200F, B-3001 Leuven, Belgium}
\date{\today}

\begin{abstract}
The superexchange theory predicts dominant antiferromagnetic kinetic interaction when the orbitals accommodating magnetic electrons are covalently bonded through diamagnetic bridging atoms or groups. 
Here we show that explicit consideration of magnetic and (leading) bridging orbitals, together with the electron transfer between the former, reveals a strong ferromagnetic kinetic exchange contribution.
First principle calculations show that it is comparable in strength with antiferromagnetic superexchange in a number of magnetic materials with diamagnetic metal bridges. 
In particular, it is responsible for a very large ferromagnetic coupling ($-10$ meV) between the iron ions in a Fe$^{3+}$-Co$^{3+}$-Fe$^{3+}$ complex. 
Furthermore, we find that the ferromagnetic exchange interaction turns into antiferromagnetic by substituting the diamagnetic bridge with magnetic one. The phenomenology is observed in two series of materials, supporting the significance of the ferromagnetic kinetic exchange mechanism.
\end{abstract}

\maketitle

\section{Introduction}
Anderson's superexchange theory \cite{Anderson1959} plays a central role in the description of exchange interactions in correlated magnetic insulators. 
It provides in particular an explanation of phenomenological Goodenough-Kanamori rules \cite{Goodenough1958, Goodenough1963, Kanamori1959}.
This theory identifies the orbitals at which reside the unpaired (magnetic) electrons - the Anderson magnetic orbitals (AMOs) - via a minimization of electron repulsion on magnetic sites. 
For non-negligible electron transfer ($b$) between these magnetic orbitals, the theory predicts strong kinetic antiferromagnetic interaction between localized spins, $J = 4b^2/U$, where $U$ is the electron repulsion on magnetic sites.
When $b$ is suppressed e.g., for symmetry reasons \cite{Kanamori1959, Kahn1993}, weaker ferromagnetic interactions of non-kinetic origin, such as, potential exchange \cite{Anderson1959, Anderson1963}, Goodenough's mechanism \cite{Goodenough1958, Goodenough1963} and spin-polarization (the RKKY mechanism) \cite{RKKY1,RKKY2,RKKY3} become dominant. 

Various developments of this theory have been proposed in the last decades \cite{Anderson1963, VanVleck1962, Gondaira1966, Ginsberg1971, Hay1975, Geertsma1990, Khomskii2014, Streltsov2017}. 
Moreover, the AMOs have been used in the analysis of exchange interactions derived from first-principles calculations \cite{Kahn1993, Loth1981, Caballol1997, Calzado2002}.
The physics of Anderson's model lies on the basis of the derivation of exchange parameters through spin-unrestricted broken-symmetry density functional theory (DFT) widely employed nowadays \cite{Noodleman1981, Soda2000, Ruiz2004, Neese2009, Riedl2019}. 
The superexchange theory \cite{Anderson1959, Anderson1963} has been extended to treat exchange interactions between orbitally degenerate sites \cite{Kugel1972, Fuchikami1978, Kugel1982, Tokura2000, Khaliullin2005},
in the presence of spin-orbit coupling on the metal ions \cite{Moriya1960, Elliott, Hartmann-Boutron, Mironov2003, Jackeli2009, Santini2009, Iwahara2015, Iwahara2016},
and beyond the second order perturbation theory after $b$, leading to biquadratic \cite{Anderson1963, Moriya1960, Harris1963, Huang1964}
and ring \cite{Takahashi1977, MacDonald1988, Roger1989} exchange interactions. 

A different extension of the theory was proposed by Geertsma \cite{Geertsma}, Larson {\it et al} \cite{Larson1985}, and Zaanen and Sawatzky \cite{Zaanen1987}
through explicit consideration of the orbitals of bridging diamagnetic atoms or groups along with the orbitals accommodating the magnetic electrons. 
Such an extension allowed for a concomitant description of high-energy excitations and exchange interaction in charge-transfer insulators \cite{Zaanen1985}.
Another reason for this extension was the claim that Anderson's theory would break down when the ligand-to-metal electron transfer energy becomes lower than the metal-to-metal electron transfer energy \cite{Zaanen1987}.
However, a detailed analysis has shown that the predictions of this extended model for the low-lying states are basically the same as of Anderson's model when only metal-ligand electron transfer is taken into account \cite{vandenHeuvel2007}.
The situation changes crucially when the metal-to-metal electron transfer is added to the model. 
In this case a strong ferromagnetic contribution of kinetic origin can arise \cite{Tasaki1995, Chibotaru1996, Chibotaru2003}.
Despite the fact that this mechanism has been mentioned on different occasions \cite{Tasaki1995, Penc1996, Tasaki1998, Tasaki2003, Tamura2019, Tasaki2020},
its relevance to existing materials has not been clarified. 

In this work, we elucidate the conditions for strong ferromagnetic kinetic exchange interaction. 
Combining model description with first-principles calculations, we prove the importance of this exchange mechanism in ferromagnetic metal compounds and its dominant contribution in cases of very strong ferromagnetic coupling between distant metal sites. 
We show that also in materials not exhibiting (strong) ferromagnetism, the kinetic ferromagnetic contribution is crucial for the annihilation of the antiferromagnetic superexchange.

\section{Basic three-site model}
\label{Sec:Basicmodel}

\begin{figure}[tb]
\includegraphics[width=0.6\linewidth]{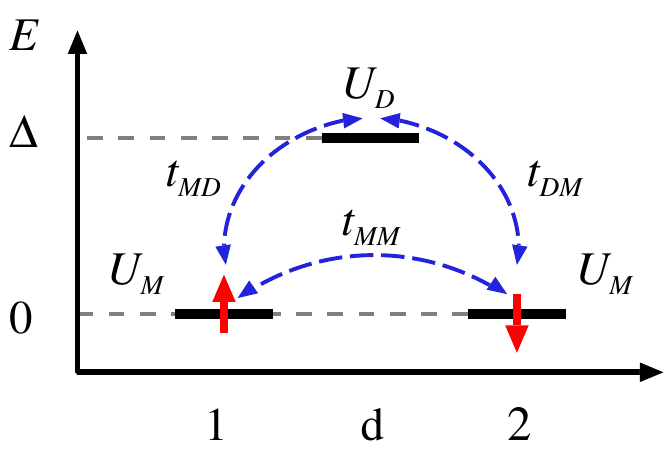}
\caption{Basic three-site model for the system consisting of two paramagnetic (1,2) and a bridging diamagnetic ($d$) sites.
The parameters correspond to Eq. (\ref{Eq:H}).}
\label{Fig:basic_model}
\end{figure}

\subsection{Model Hamiltonian}
In a first step, we derive the AMOs as minimizing the electron repulsion between magnetic electrons in a spin-restricted broken-symmetry band (molecular) orbital picture \cite{Anderson1959, Anderson1963}.
Then we identify the common ligand orbitals in the composition of neighbor AMOs and approximate them by Wannier transformation of a group of suitable band (molecular) orbitals. 
The resulting localized bridging orbitals (LBOs) mainly reside at the diamagnetic atom or group bridging the neighbor paramagnetic sites. 
Extracting these orbitals from the AMOs via an orthogonal transformation, we end up with localized magnetic orbitals (LMO), which are more localized on the paramagnetic sites than the corresponding AMOs but now strongly overlap with neighbor LBOs. 
The exchange interaction is derived from a many-body treatment of electrons in LMOs of two chosen paramagnetic sites and LBOs of the bridging diamagnetic atom or group.  

We first consider the simplest model involving only two LMOs and one LBO (Fig. \ref{Fig:basic_model}), 
\begin{eqnarray}
 \hat{H} &=& \sum_{\sigma = \uparrow, \downarrow} 
 \left[ 
  \Delta \hat{n}_{d\sigma} 
 + t_{MM} (\hat{a}_{1\sigma}^\dagger \hat{a}_{2\sigma} + \hat{a}_{2\sigma}^\dagger \hat{a}_{1\sigma})
 \right.
\nonumber\\
 &&+ 
 \left.
 t_{MD}
  (\hat{a}_{1\sigma}^\dagger \hat{a}_{d\sigma} + \hat{a}_{d\sigma}^\dagger \hat{a}_{1\sigma})
 +
 t_{DM}
  (\hat{a}_{2\sigma}^\dagger \hat{a}_{d\sigma} + \hat{a}_{d\sigma}^\dagger \hat{a}_{2\sigma})
 \right]
\nonumber\\
 &&
 + U_M (\hat{n}_{1\uparrow} \hat{n}_{1\downarrow} + \hat{n}_{2\uparrow} \hat{n}_{2\downarrow} )
 + U_D \hat{n}_{d\uparrow} \hat{n}_{d\downarrow}.
\label{Eq:H}
\end{eqnarray}
Here, $1,2$ and $d$ indicate the paramagnetic and the diamagnetic sites, respectively, 
$\sigma$ $(= \uparrow, \downarrow)$ is electron spin projection,
$\hat{a}_{i\sigma}^\dagger$ ($\hat{a}_{i\sigma}$) is the electron creation (annihilation) operator in the localized orbital on the sites, 
$i$ $(= 1, 2, d)$ with $\sigma$, $\hat{n}_{i\sigma} = \hat{a}_{i\sigma}^\dagger \hat{a}_{i\sigma}$,
$t_{MD/DM}$ and $t_{MM}$ are the corresponding electron transfer parameters, 
$\Delta > 0$ is the gap between the diamagnetic and paramagnetic orbital level, and $U_M$ and $U_D$ are on-site Coulomb repulsion parameters within the LMO and LBO, respectively. 
For symmetric magnetic sites considered below, the following relations hold: $t_{MD} = t_{DM}$ or $t_{MD} = -t_{DM}$.

\begin{figure*}[bt]
\begin{tabular}{lll}
(a) &~ (b) &~ (c) 
\\
\includegraphics[height=4.5cm]{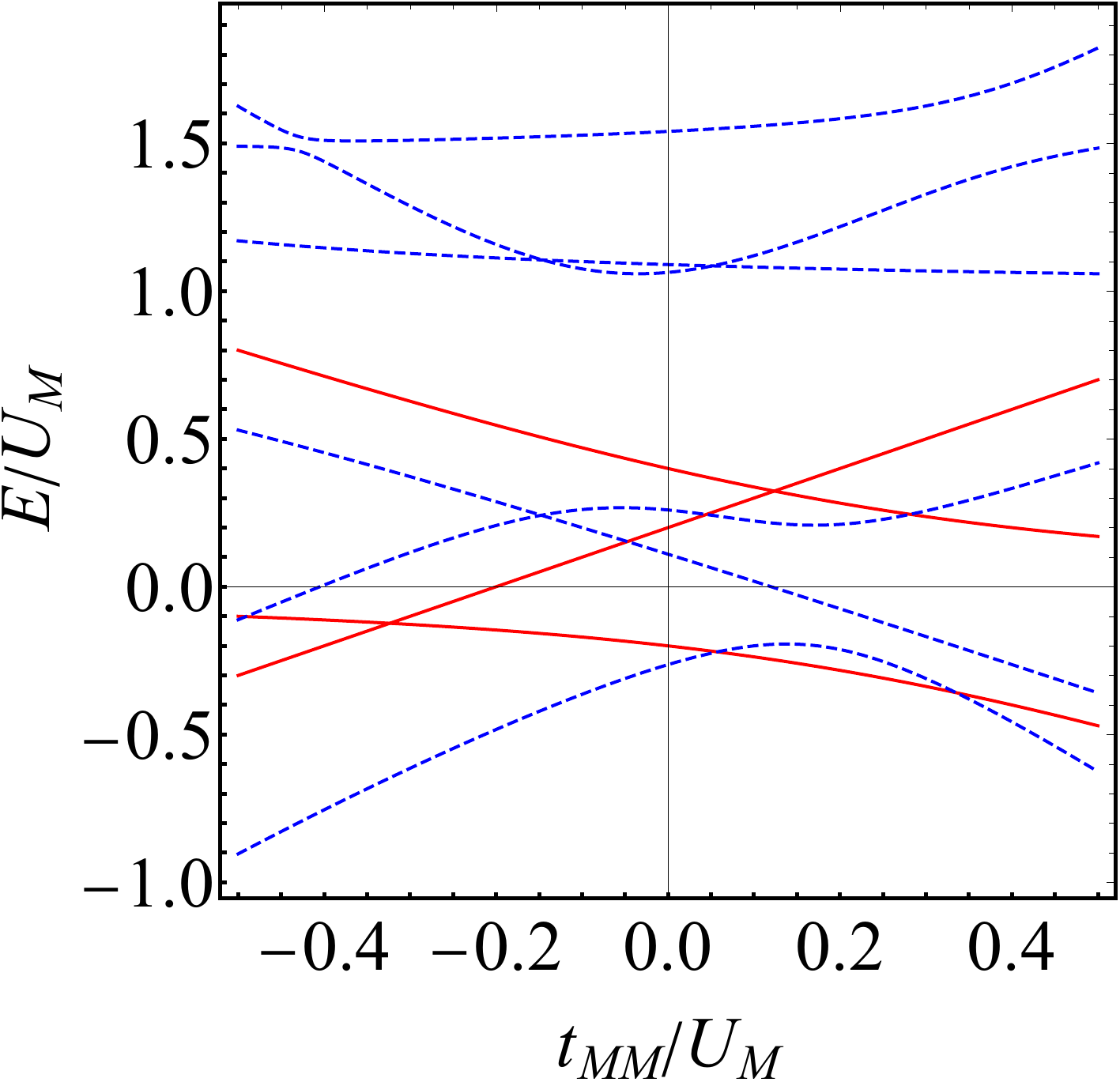} 
&
~
\includegraphics[height=4.5cm]{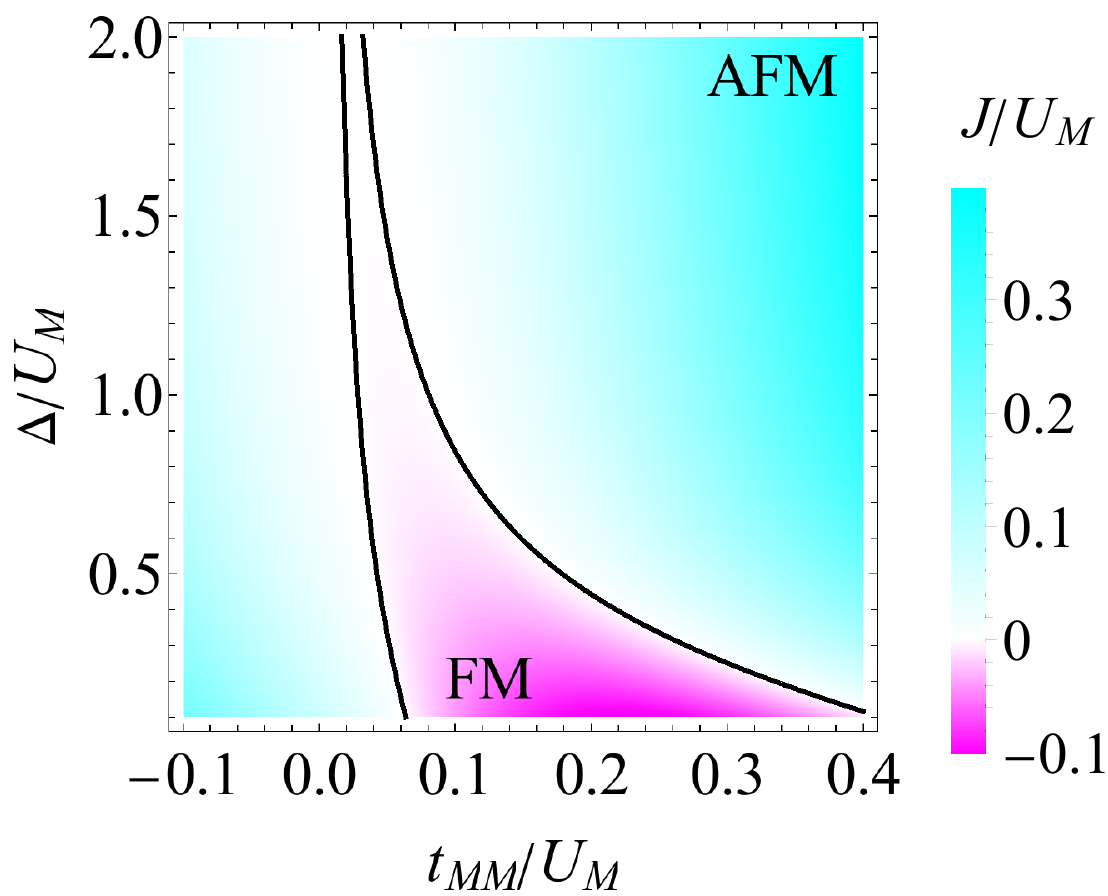} 
&
~
\raisebox{.125\height}{
\includegraphics[height=3.5cm]{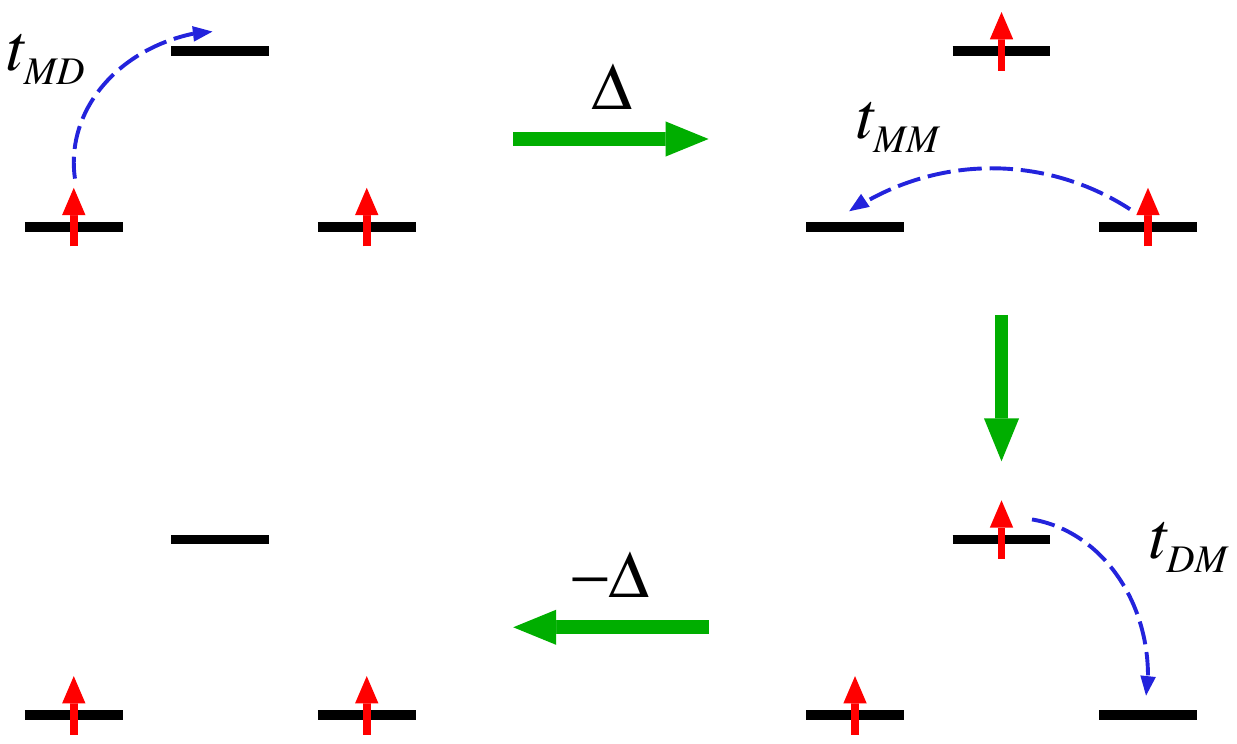}
}
\hfill
\end{tabular}
\caption{
(a) Energy levels diagram of the three-site model (\ref{Eq:H}) for $t_{MD} = t_{DM}$, $t_{MD}/U_M = \Delta/U_M = 0.2$ and $U_D/U_M = 1$.
The solid red and dashed blue lines indicate triplet and singlet states, respectively.  
(b) Exchange parameter diagram [other parameters than indicated on the axes are the same as in (a)].
$J$ is for Eq. (\ref{Eq:HHeisenberg}).
(c) Third-order process responsible for ferromagnetic kinetic exchange contribution. 
}
\label{Fig:model}
\end{figure*}

The model (\ref{Eq:H}) always reduces to two unpaired particles localized at the LMOs, which are electrons when the LBO on the diamagnetic site is empty and holes when this is doubly occupied. 
In the latter case, all one-electron parameters in Eq. (\ref{Eq:H}) change the sign except for $\Delta$ which becomes $-\Delta + 2U_D$ \cite{Chibotaru2003}, remaining always positive in magnetic insulators. 
For $t_{MM} = 0$, the Hamiltonian (\ref{Eq:H}) reduces to the earlier considered 3-orbital model \cite{Geertsma, Larson1985, Zaanen1987}. 
We stress, however, that this limit is often unrealistic because the LMOs and the LBO are not atomic orbitals but instead have ``tails'' which extend on neighbor sites, in analogy with AMOs \cite{Anderson1959, Anderson1963}.

\subsection{Ferromagnetic kinetic exchange interaction}
The calculated energy spectrum of model (\ref{Eq:H}) is shown in Fig. \ref{Fig:model}(a). 
One can see that the system exhibits strong ferromagnetism for relatively large values of $t_{MM}$, further enhanced for small $\Delta$ [Fig. \ref{Fig:model}(b)].
We emphasize that it arises without Hund's rule coupling and potential exchange interaction, which are not included in Eq. (\ref{Eq:H}).
To unravel the mechanism of the ferromagnetism, we consider $|t_{MD}|,|t_{MM}|\ll U_M, U_D, |\Delta|$, 
and obtain in the fourth order of perturbation theory the expression for the exchange parameter $J$ for the spin $1/2$ Heisenberg model,
\begin{eqnarray}
\hat{H}_\text{ex} &=& J \hat{\bm{s}}_1 \cdot \hat{\bm{s}}_2.
\label{Eq:HHeisenberg}
\end{eqnarray}
The ground energies for the ferro- and antiferromagnetic states are calculated as 
\begin{eqnarray}
 E_\text{F} &=&
  -\frac{2t_{MD}^2}{\Delta} + \frac{4t_{MD}^4}{\Delta^3} 
  - \frac{2t_{MD}t_{DM}t_{MM}}{\Delta^2}, 
\label{Eq:EF}
\\
 E_\text{AF} &=& 
 -\frac{2t_{MD}^2}{\Delta} + \frac{4t_{MD}^4}{\Delta^3} 
 - \frac{4}{U_M} \left(t_{MM} - \frac{t_{MD}t_{DM}}{\Delta}\right)^2
\nonumber\\
 &&
 - \frac{8t_{MD}^2t_{DM}^2}{\Delta^2(U_D + 2\Delta)}
 + \frac{2t_{MD}t_{DM}t_{MM}}{\Delta^2}
 + \frac{16t_{MM}^4}{U_M^3},
\label{Eq:EAF}
\end{eqnarray}
respectively
\footnote{
For the calculations of higher order terms, the approach using resolvent is convenient (see e.g. Ref. \cite{Takahashi1977}).
}.
The energy gap between them, $E_\text{F} - E_\text{AF}$, corresponds to the exchange parameter $J$ (\ref{Eq:HHeisenberg}):
\begin{eqnarray}
 J &=& 
 \frac{4}{U_M} \left(t_{MM} - \frac{t_{MD}t_{DM}}{\Delta} \right)^2 
 + \frac{8t_{MD}^2t_{DM}^2}{\Delta^2(U_D + 2\Delta)}
\nonumber\\
 &&
  - \frac{4t_{MD}t_{DM}t_{MM}}{\Delta^2}
  - \frac{16t_{MM}^4}{U_M^3}.
\label{Eq:J}
\end{eqnarray}
The first and second terms in Eq. (\ref{Eq:J}), K1 and K2, are always antiferromagnetic ($>0$), and the fourth term (K4) is ferromagnetic $(<0)$.  
The third term (K3) becomes ferromagnetic for $t_{MD}t_{DM}t_{MM} > 0$ and is antiferromagnetic otherwise. 
According to the order of the perturbation, the first and the third terms are dominant, and the nature of $J$ is mainly determined by their competition.

The ferromagnetic contribution K3 originates from cyclic electron transfer processes avoiding double occupation of any of three orbitals [Fig. \ref{Fig:model}(c)].
It can be called the {\it ferromagnetic kinetic exchange interaction}.
Note that the contribution of this mechanism to the energy of the ferromagnetic state, $-2t_{MD}t_{DM}t_{MM}/\Delta^2$ [the factor 2 is due to a cyclic processes, similar to Fig. \ref{Fig:model}(c) but in anticlockwise sense], is opposite to the case of antiferromagnetic state, because of the sign change in the latter [see the third and fifth terms in Eqs. (\ref{Eq:EF}) and (\ref{Eq:EAF}), respectively].

\subsection{Ferromagnetism within Anderson's approach}
It should be noted that the ferromagnetic kinetic exchange contribution (K3) is not fully captured by Anderson's approach \cite{Anderson1959}. 
Projecting the basic three-site model, Eq. (\ref{Eq:H}), on the space of two AMOs, we obtain a tight-binding model (see for derivation Appendix \ref{A:Anderson}):
\begin{eqnarray}
 \hat{H} &=&
 E_\text{HF} + \hat{H}_A 
 + \sum_{\sigma = \uparrow, \downarrow}
 b' \left(\hat{N}_{1,-\sigma} + \hat{N}_{2,-\sigma} \right) 
\nonumber\\
 && \times
  \left(\hat{A}_{1\sigma}^\dagger \hat{A}_{2\sigma} + \hat{A}_{2\sigma}^\dagger \hat{A}_{1\sigma} \right) 
 + U' \left[ \hat{N}_{1\uparrow} \hat{N}_{2\downarrow} + \hat{N}_{2\uparrow} \hat{N}_{1\downarrow} 
 \right.
\nonumber\\
 && + 
 \left.
  \left(\hat{A}_{1\uparrow}^\dagger \hat{A}_{2\uparrow} + \hat{A}_{2\uparrow}^\dagger \hat{A}_{1\uparrow} \right)
  \left(\hat{A}_{1\downarrow}^\dagger \hat{A}_{2\downarrow} + \hat{A}_{2\downarrow}^\dagger \hat{A}_{1\downarrow} \right)
 \right],
\label{Eq:HAMOgen}
\\
 \hat{H}_A &=& 
 \sum_{\sigma} 
 \tau b \left(\hat{A}_{1\sigma}^\dagger \hat{A}_{2\sigma} + \hat{A}_{2\sigma}^\dagger \hat{A}_{1\sigma}\right)
 + \sum_{i=1,2} U \hat{N}_{i\uparrow} \hat{N}_{i\downarrow},
\label{Eq:HAMO}
\end{eqnarray}
where $E_\text{HF}$ is the restricted open shell Hartree-Fock energy for the ferromagnetic state, 
$\hat{A}_{i\sigma}$ ($\hat{A}_{i\sigma}^\dagger$) is the electron annihilation (creation) operator in the AMO centered at site $i$, 
$\hat{N}_{i\sigma} = \hat{A}_{i\sigma}^\dagger \hat{A}_{i\sigma}$,
$\tau = t_{MD}/t_{DM}$, 
$b$ is the effective electron transfer parameter between the two AMOs (\ref{Eq:bAMO}),
$U$ is the energy of electron promotion between AMOs (\ref{Eq:UAMO}), 
$U'$ is the intersite potential exchange interaction parameter (\ref{Eq:U'}), 
and $b'$ is the Coulomb repulsion assisted transfer parameter (\ref{Eq:b'}). 
Although the last term in Eq. (\ref{Eq:HAMOgen}) is different from the standard potential exchange interaction (\ref{Eq:HPE}), we use the name because the resulting spin dependent shift of energy levels resembles it.

\begin{figure}[tb]
\includegraphics[width=0.7\linewidth]{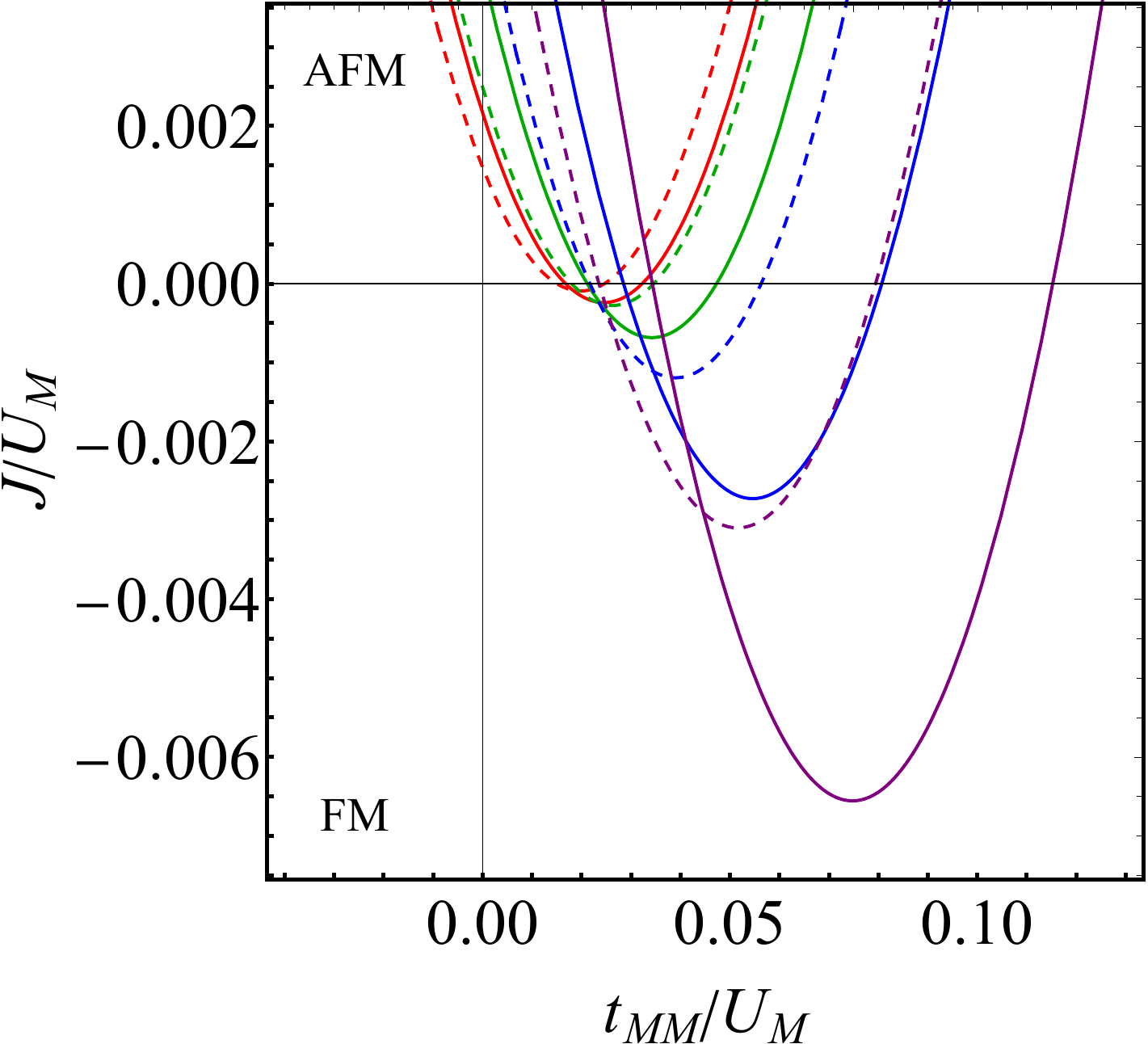}
\caption{
Exchange parameters calculated by exact diagonalization (solid) and within Anderson's model (dashed).
The red, green, blue, and purple lines indicate $\Delta/U_M = 2, 1.5, 1, 0.75$, respectively.
$t_{MD}/U_M=t_{DM}/U_M=0.2$ and $U_D/U_M = 1$ are used in all calculations.
}
\label{Fig:J_exact_Anderson}
\end{figure}

The second Hamiltonian (\ref{Eq:HAMO}) is regarded as the original Anderson tight-binding model (or Hubbard model) \cite{Anderson1959}. 
As is well known, the exchange parameter $J$ from the model (\ref{Eq:HAMO}) gives the antiferromagnetic contribution.
The extended model (\ref{Eq:HAMOgen}) retains all the interaction terms appearing in the approach based on the AMOs, and hence, describes the magnetic properties more accurately (see for detailed analysis of Eq. (\ref{Eq:H}) without next-nearest neighbor transfer, $t_{MM} = 0$, Ref. \cite{vandenHeuvel2007}). 
The calculated $J$ within this approach (dashed line) is shown in Fig. \ref{Fig:J_exact_Anderson} in comparison with the exact treatment (solid line).
Indeed, a finite $t_{MM}$ merely modifies the effective transfer parameter $b$ between the AMOs, i.e., the antiferromagnetic kinetic exchange: 
For $t_{MM} > 0$, the ferromagnetic kinetic exchange contribution [K3 in Eq. (\ref{Eq:J})] is only partly recovered within the extended Anderson's theory through mainly the potential exchange contribution as shown below. 
However, it is much underestimated compared to the exact treatment (Fig. \ref{Fig:J_exact_Anderson}).

For further insight, the exchange interaction parameter is calculated within perturbation theory in the case of $|t_{MD}|,|t_{MM}|\ll U$. 
The energy level for the ferromagnetic state is $E_\text{HF}$, and its expression becomes the same as Eq. (\ref{Eq:EF}) (see for calculations Appendix \ref{A:Anderson}).
This coincidence is a consequence that the ferromagnetic ground state (with maximum spin projection) is described exactly within single Slater determinant.
On the other hand, the antiferromagnetic ground state energy (\ref{Eq:ELSAnderson}) differs much from Eq. (\ref{Eq:EAF}). 
The leading terms of the exchange parameter $J$ are
\begin{eqnarray}
 J &=& -2U' + \frac{4(b+b')^2}{U-U'} - \frac{16(b+b')^4}{(U-U')^3}.
\label{Eq:JAMO_approx}
\end{eqnarray}
The second of Eq. (\ref{Eq:JAMO_approx}) gives the strong antiferromagnetic contribution. 
The ferromagnetic contribution arises from the potential exchange term (the first term) and a contribution of K4 type (the third term).

\subsection{Condition for strong ferromagnetism}
The necessary condition for a dominant ferromagnetic kinetic contribution is the right sign and a large value of $t_{MM}$. 
The existence of non-negligible $t_{MM}$ is expected for LMOs extending on neighbor paramagnetic sites. 
This occurs when the relevant bands (molecular orbitals) involve several atomic orbitals centered on different atoms in the unit cell (molecule). 
Then, the corresponding Wannier orbitals will not be completely localized, leading to non-negligible overlap between neighbor LMOs. 
In an opposite situation, when the common bridging orbitals in the composition of neighbor AMOs are contained in the same number of relevant bands (molecular orbitals), the Wannier transformation of the latter will result in LMOs almost coinciding with atomic orbitals and LBOs well localized on the bridging diamagnetic groups. 
An example is a family of superconducting cuprates, in which the low-energy states are described by a three-orbital model for the CuO$_2$ plane \cite{Emery1987}, involving the almost net atomic $3d_{x^2-y^2}$ orbital on Cu and $2p_x$ ($2p_y$) orbitals on O. 
The latter lead to small $t_{MM}$ and negligible kinetic ferromagnetic exchange contribution ($t_{MD}t_{DM}t_{MM} > 0$), which is in accord with a very large antiferromagnetic exchange interaction in cuprates \cite{Nagaosa1997}.
On the other hand, the sign of $t_{MM}$ is unambiguously determined by the type of the localized orbitals (e.g., $d$, $f$), which is not influenced by the hybridization of the Wannier orbitals.

According to Eq. (\ref{Eq:J}), the $t_{MM}$ of a right sign not only gives rise to a ferromagnetic kinetic contribution but concomitantly reduces the antiferromagnetic one. 
However, the largest ferromagnetic $J$ is not achieved at a $t_{MM}$ quenching K1 but at a larger value, $t_{MM} \approx (t_{MD}t_{DM}/\Delta)(1+U_M/2\Delta)$. 
The expression in Eq. (\ref{Eq:J}) then becomes 
\begin{eqnarray}
 J_\text{ferro}^\text{max} &\approx& -4\frac{(t_{MD}t_{DM}/\Delta)^2}{\Delta} \left( \frac{U_D}{U_D+2\Delta} + \frac{U_M}{4\Delta} \right).
\label{Eq:Jferro}
\end{eqnarray}
Counter-intuitively, the ferromagnetic coupling increases linearly with $U_M$.
Besides, it rises very fast with diminishing $\Delta$, a feature also confirmed by non-perturbative treatment [Fig. \ref{Fig:model}(b)]. 
Small $\Delta$ (strong metal-ligand hybridization) is expected in late transition metal compounds, which are thus primary candidates for the observation of strong kinetic ferromagnetism.

\subsection{Switching of ferromagnetic kinetic mechanism}
\label{Sec:switch}
\begin{figure}[tb]
 \includegraphics[height=4.5cm]{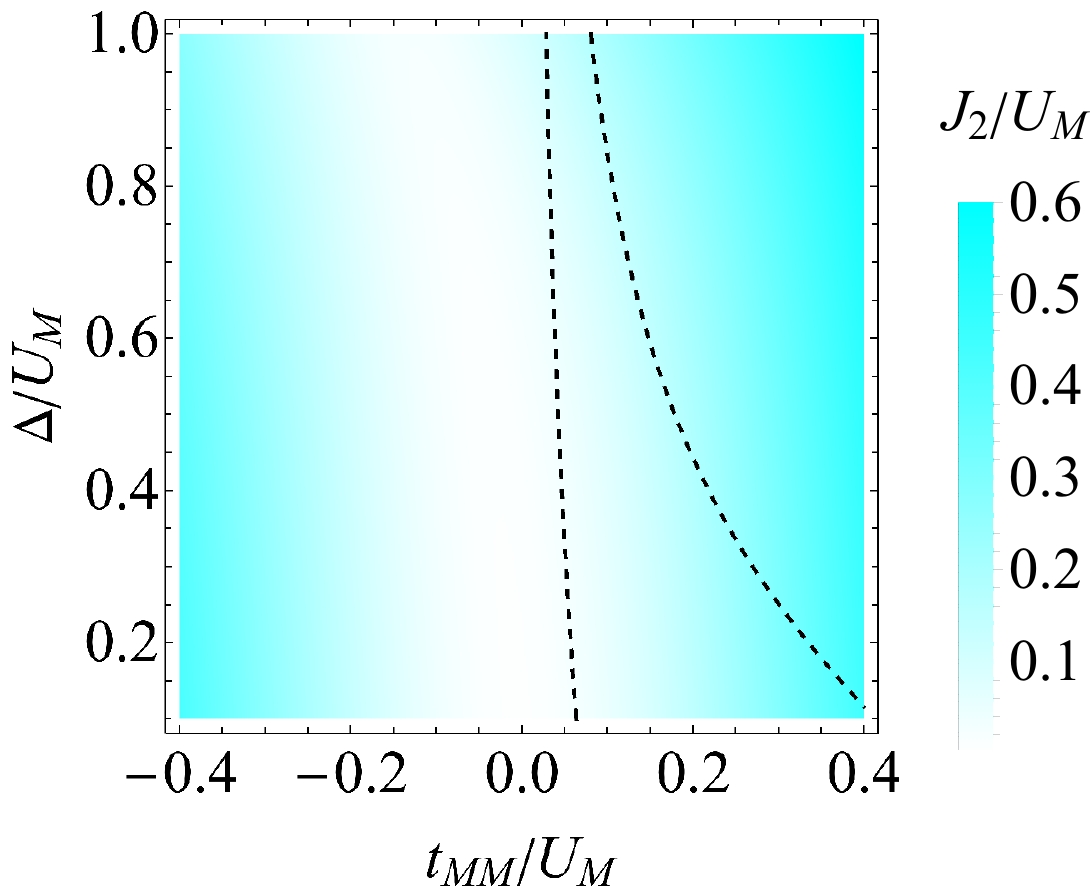} 
\caption{$J_2$ diagram. The parameters are the same as in Fig. \ref{Fig:model}(b). The dashed lines outline the ferromagnetic domain in the case of diamagnetic bridge [Fig. \ref{Fig:model}(b)]. }
\label{Fig:J2}
\end{figure}

A similar treatment shows that adding one electron or hole to the empty or doubly occupied LBO turns the initially dominant ferromagnetic kinetic interaction into an antiferromagnetic one of comparable strength (see Fig. \ref{Fig:J2}). 
The Heisenberg model for the electron or hole doped system is written as 
\footnote{
The energy level for the ferromagnetic state is $J_1/2 + J_2/4$, 
and those for the antiferromagnetic states are $-J_1 + J_2/4$ and $-3J_2/4$. 
The eigenstates for the latter (spin projection +1/2) are 
\unexpanded{$(|\downarrow \uparrow \uparrow\rangle -2 |\uparrow \downarrow \uparrow\rangle + |\uparrow \uparrow \downarrow\rangle)/\sqrt{6}$}
and 
\unexpanded{$(-|\downarrow \uparrow \uparrow\rangle + |\uparrow \uparrow \downarrow\rangle)/\sqrt{2}$},
where 
\unexpanded{$|\sigma_1 \sigma_d \sigma_2\rangle$} 
\unexpanded{$(\sigma_i = \uparrow, \downarrow)$}
are the spin configurations of the system. 
}
\begin{eqnarray}
 \hat{H} &=& J_1 (\hat{\bm{s}}_1 \cdot \hat{\bm{s}}_d + \hat{\bm{s}}_1 \cdot \hat{\bm{s}}_d) + J_2 \hat{\bm{s}}_1 \cdot \hat{\bm{s}}_2.
\label{Eq:H_doped}
\end{eqnarray}
Within the perturbation theory ($|t_{MM}|, |t_{MD/DM}| \ll U_M, U_D, U_M - \Delta$), the exchange parameters $J_1$ and $J_2$ are calculated as 
\begin{eqnarray}
 J_1 &=& 
 \frac{2t_{MD}^2}{U_M-\Delta+\tau t_{MM}} + \frac{2t_{MD}^2}{U_D+\Delta-\tau t_{MM}},
\label{Eq:J1_doped}
\\
 J_2 &=& \frac{4t_{MM}^2}{U_M}
 + \frac{2t_{MD}t_{DM}t_{MM}}{(U_M-\Delta)^2} 
 + \frac{4t_{MD}t_{DM}t_{MM}}{U_M(U_M-\Delta)}
\nonumber\\
 &&-
   \frac{2t_{MD}t_{DM}t_{MM}}{(U_D+\Delta)^2} 
 - \frac{4t_{MD}t_{DM}t_{MM}}{U_M(U_D+\Delta)}.
\label{Eq:J2_doped}
\end{eqnarray}
$J_1$ is antiferromagnetic because $\Delta$ is smaller than $U_M$: otherwise the added electron/hole occupies the paramagnetic sites rather than the diamagnetic site. 
The first term of $J_2$ is antiferromagnetic and the remaining terms become both antiferromagnetic and ferromagnetic depending on the sign of $t_{MD}t_{DM}t_{MM}$. 
When $t_{MD}t_{DM}t_{MM} > 0$, the last two terms of $J_2$ (\ref{Eq:J2_doped}) become ferromagnetic as K3 (\ref{Eq:J}), while the ferromagnetism is quenched due to the following reasons. 
First, contrary to the undoped system (\ref{Eq:EF}), the ferromagnetic state is not stabilized by hybridization because the cyclic electron transfer processes [Fig. \ref{Fig:model}(c)] are forbidden by Pauli's exclusion principle.
Second, the denominators of the ferromagnetic terms of $J_2$ tend to be large in comparison with the K3 term, resulting in weaker ferromagnetic contribution than the K3.
Finally, these contributions are canceled by the antiferromagnetic contributions (the second and the third terms as well as the first term of $J_2$). 
In particular, the second and the third terms of $J_2$ are large because of the partial cancellation of $U_M$ and $\Delta$, which gradually increases as $U_M - \Delta$ decreases ($\Delta/U_M$ increases) as seen in our numerical analysis (Fig. \ref{Fig:J2}).  
Therefore, the kinetic ferromagnetic mechanism for next-nearest neighbor exchange pairs is quenched by adding an electron or hole to the LBO.

On the other hand, in the case of $t_{MD}t_{DM}t_{MM} < 0$, the second and the third terms in $J_2$ (\ref{Eq:J2_doped}) are ferromagnetic.
These ferromagnetic contributions become comparable to the strongest antiferromagnetic contribution (the first term) when $t_{MM}$ is smaller than $t_{MD}t_{DM}/(U_M - \Delta)$. 
Indeed, our numerical calculations show that $J_2$ becomes very weak antiferromagnetic for $-t_{MD}t_{DM}/(U_M - \Delta) \alt t_{MM} < 0$ (Fig. \ref{Fig:J2}).

\section{Relevance of the ferromagnetic kinetic exchange mechanism to magnetic  materials}
\subsection{First-principles based approach}
The ferromagnetic kinetic exchange mechanism is further investigated in several magnetic materials with diamagnetic metal bridges.
Examples considered include complexes Fe$^{3+}$-Co$^{3+}$-Fe$^{3+}$ \cite{Glaser1999}, and Cu$^{2+}$-Cr$^{6+}$-Cu$^{2+}$ and Cu$^{2+}$-Mo$^{6+}$-Cu$^{2+}$ \cite{Oshio1996}, and a quasi-one-dimensional Cu chain in La$_4$Ba$_2$Cu$_2$O$_{10}$ \cite{Mizuno1990, Masuda1991}.
In these systems, Fe$^{3+}$ ($d^5$) and Cu$^{2+}$ ($d^9$) ions are the magnetic ions with $s=1/2$, while Co$^{3+}$ ($d^6$), Cr$^{6+}$ ($d^0$), Mo$^{6+}$ ($d^0$) and La$^{3+}$ belong to diamagnetic bridges.
Despite the large distance between paramagnetic centers, they all (except Cu-Mo-Cu) display ferromagnetic exchange interaction. 

In order to achieve a realistic description of exchange contributions, the results of first-principles calculations were mapped into an extended three-sites model 
which, contrary to the basic model in Eq. (\ref{Eq:H}), includes all relevant LBOs on the diamagnetic bridging site and the Coulomb and potential exchange interactions between the LMOs and LBOs:
\begin{eqnarray}
 \hat{H} &=& \hat{H}_0 + \hat{H}_\text{t} + \hat{H}_\text{Coul} + \hat{H}_\text{PE},
\label{Eq:Hreal}
\\
 \hat{H}_0 &=& \sum_{d} \sum_{\sigma = \uparrow \downarrow} \Delta_d \hat{n}_{d\sigma},
\label{Eq:H0}
\\ 
 \hat{H}_t &=& \sum_{i = 1,2} \sum_{d} \sum_{\sigma = \uparrow \downarrow}
\left(
  t_{Md} 
   \hat{a}_{i\sigma}^\dagger \hat{a}_{d\sigma} 
+ t_{dM} \hat{a}_{d\sigma}^\dagger \hat{a}_{i\sigma} 
 \right)
\nonumber\\
 && + 
 \sum_{\sigma = \uparrow \downarrow}
  t_{MM} \left(
   \hat{a}_{1\sigma}^\dagger \hat{a}_{2\sigma} + \hat{a}_{2\sigma}^\dagger \hat{a}_{1\sigma} 
 \right)
\nonumber\\
 &&+ 
 \sum_{d<d'} \sum_{\sigma = \uparrow \downarrow} 
 t_{dd'} 
 (\hat{a}_{d\sigma}^\dagger \hat{a}_{d'\sigma} 
 + 
 \hat{a}_{d'\sigma}^\dagger \hat{a}_{d\sigma})
\label{Eq:Ht}
\\
 \hat{H}_\text{Coul}
  &=& \sum_{i = 1,2} U_M \hat{n}_{i\uparrow} \hat{n}_{i\downarrow}
      + \sum_{d} U_d \hat{n}_{d\uparrow} \hat{n}_{d\downarrow}
      + V_{MM} \hat{n}_{1} \hat{n}_{2}
\nonumber\\
   &&+ \sum_{i=1,2} \sum_{d} V_{Md} \hat{n}_i \hat{n}_d
   + \sum_{d<d'} V_{dd'} \hat{n}_{d} \hat{n}_{d'},
\label{Eq:HCoul}
\\
 \hat{H}_\text{PE} 
  &=& \sum_{i = 1,2} \sum_{d} \sum_{\sigma \sigma' = \uparrow, \downarrow} J_{Md} \hat{a}_{i\sigma}^\dagger \hat{a}_{d\sigma'}^\dagger \hat{a}_{i\sigma'} \hat{a}_{d\sigma}
\nonumber\\
   &&+ \sum_{\sigma \sigma' = \uparrow, \downarrow} J_{MM} \hat{a}_{1\sigma}^\dagger \hat{a}_{2\sigma'}^\dagger \hat{a}_{1\sigma'} \hat{a}_{2\sigma}.
\label{Eq:HPE}
\end{eqnarray}
Here $d$ indicates the LBO on the bridging diamagnetic site,
$\Delta_d$ is the energy gap between the LBO $d$ and the LMO levels,
$t_{id}$ are the electron transfer parameter between the corresponding orbitals,
$V_{Md}$ the intersite Coulomb repulsion,
$V_{dd'}$ is the Coulomb repulsion between different LBOs,
$J_{MM}$ and $J_{Md}$ the potential exchange parameters,
and $\hat{n}_{j} = \sum_{\sigma = \uparrow, \downarrow} \hat{n}_{j\sigma}$ $(j = 1,2,d)$.
As in the case of the basic three-site model, $t_{Md}$ and $t_{dM}$ fulfill either $t_{Md} = t_{dM}$ or $t_{Md} = -t_{dM}$.

The energy eigenstates of the Hamiltonian are derived in two ways: direct numerical diagonalization and perturbation theory.
In the former case, the Hamiltonian matrix for the three center complexes or fragment is built using all electron configurations constructed with LMOs and LBOs as the basis and DFT parameters, and then numerically diagonalized. 
Within the fourth order perturbation theory, the contributions to the Heisenberg exchange parameter are calculated as 
\begin{eqnarray}
 J &=& J_{\text{K}1} + J_{\text{K}2} + J_{\text{K}3} + J_{\text{K}4} + J_{\text{PE}},
\\
 J_{\text{K}1} &=& 
 \frac{4}{U_M-V_{MM}} \left(t_{MM} - \sum_d \frac{t_{Md}t_{dM}}{\Delta_d - V_{MM} + V_{Md}}\right)^2,
\nonumber\\
\label{Eq:JK1}
\\
 J_{\text{K}2} &=& 
  \sum_{dd'} 
  \frac{2t_{Md}t_{dM}t_{Md'}t_{d'M}}{\Delta_d+\Delta_{d'}-V_{MM}+V_{dd'}}
\nonumber\\
 && \times
  \left(\frac{1}{\Delta_d-V_{MM}+V_{Md}}+\frac{1}{\Delta_{d'}-V_{MM}+V_{Md'}}\right)^2
\nonumber\\
 &&+
  \sum_{d<d'} 
  \frac{4t_{dd'}t_{MM}(t_{Md}t_{d'M} + t_{Md'}t_{dM})}{(\Delta_d -V_{MM}+V_{Md})(\Delta_{d'}-V_{MM}+V_{Md'})}
\nonumber\\
 && \times
  \frac{2}{U_M-V_{MM}}
\nonumber\\
 &&+
  \sum_{d<d'} 
  \frac{4t_{dd'}t_{MM}(t_{Md}t_{d'M} + t_{Md'}t_{dM})}{(\Delta_d -V_{MM}+V_{Md})(\Delta_{d'}-V_{MM}+V_{Md'})}
\nonumber\\
 && \times
 \left(
  \frac{1}{\Delta_d -V_{MM}+V_{Md}}
 +
  \frac{1}{\Delta_{d'} -V_{MM}+V_{Md'}}
 \right),
\nonumber\\
\label{Eq:JK2}
\\
 J_{\text{K}3} &=& -\sum_{d} \frac{4t_{Md}t_{dM}t_{MM}}{(\Delta_d - V_{MM} + V_{Md})^2},
\label{Eq:JK3}
\\
 J_{\text{K}4} &=& -\frac{16t_{MM}^4}{(U_M-V_{MM})^3}.
\label{Eq:JK4}
\\
 J_{\text{PE}} &=& -2J_{MM}.
\label{Eq:JPE}
\end{eqnarray}
The kinetic contributions (\ref{Eq:JK1})-(\ref{Eq:JK4}) for the extended model correspond to the four terms in $J$ (\ref{Eq:J}) of the basic model (\ref{Eq:H}), and are further denoted as K1-K4. 
In the last term of Eq. (\ref{Eq:JK2}), there are many cross terms involving pairs of LBOs, $t_{Md}t_{dM}t_{Md'}t_{d'M}$.
Since $t_{Md}t_{dM}$ can be both positive ($t_{Md} = t_{dM}$) and negative ($t_{Md} = -t_{dM}$), this term becomes equally positive (antiferromagnetic) and negative (ferromagnetic).

\begin{table}[tb]
\caption{
Microscopic parameters of the extended three-site model (eV).
}
\label{Table:parameter}
\begin{ruledtabular}
\begin{tabular}{ccccc}
          & Fe-Co-Fe \footnotemark[1]&  Cu-Cr-Cu &  Cu-Mo-Cu & La$_4$Ba$_2$Cu$_2$O$_{10}$ \\
\hline
$t_{MD}$  &  0.290  & $-0.499$ & $-0.554$ & 0.748 \\
$t_{MM}$  &  0.193  &   0.084  &   0.040  & 0.013 \\
$\Delta$  &  1.048\footnotemark[2]  &  3.357  &  4.774  & 6.787 \\
$\Delta'$\footnotemark[3] &  0.595  &  3.246  &  3.685  & - \\
$U_M$     &  2.912  &   4.848  &   4.482  & 3.178 \\
$U_D$     &  2.859  &   3.786  &   2.789  & 1.563 \\
$V_{MD}$  &  1.672  &   2.463  &   2.109  & 0.681 \\
$V_{MM}$  &  1.347  &   1.474  &   1.380  & 0.441 \\
$J_{MD}$\footnotemark[4]  &  0.0106 &   0.0084  &   0.0091  & - \\
$J_{MM}$\footnotemark[4]  &  0.0025 &   0.0013  &   0.0005  & $8.4 \times 10^{-5}$ \\
\end{tabular}
\end{ruledtabular}
\footnotetext[1]{ $t$ and $\Delta$ are given in hole picture. }
\footnotetext[2]{ Derived from absorption spectrum in solution. }
\footnotetext[3]{ The value allowing to reproduce the experimental $J$. }
\footnotetext[4]{ Scaled down following Ref. \cite{Ku2002} }
\end{table}

Electronic band structure calculations for all materials were performed on their experimental structure \cite{Glaser1999, Oshio1996, Ogawa1990} with the revised Perdew-Burke-Ernzerhof (PBE) functional \cite{PBEsol1} and optimized norm-conserving Vanderbilt pseudo-potentials \cite{Hamann2013}.
Using the Kohn-Sham orbitals, maximally localized Wannier functions \cite{Wannier} and one-particle interaction parameters, $t$ and $\Delta$, were derived.
Screened intra- ($U_{M/D}$) and intersite Coulomb, and potential exchange parameters were calculated within the constrained random phase approximation \cite{cRPA}.
Quantum ESPRESSO \cite{PW_2007, PW_2017} and RESPACK \cite{RESPACK_2, RESPACK_3, RESPACK_1, RESPACK_5, RESPACK_4, respack} were used for electronic structure calculations, and pseudo potentials were taken from PSEUDO DOJO \cite{VanSetten2018}, and VESTA \cite{VESTA} for plotting the orbitals.
See for details Appendix \ref{A:DFT}.

The obtained parameters of the extended three-site model for the four compounds are listed in Table \ref{Table:parameter}.
The exchange parameter $J$ of the spin-1/2 Heisenberg model (\ref{Eq:HHeisenberg}) was determined to reproduce the energy gap between the ground high- and low-spin term energies obtained by numerical diagonalization of the corresponding Hamiltonian, Eqs. (\ref{Eq:Hreal})-(\ref{Eq:HPE}).
The kinetic contributions to the exchange parameters were calculated using the corresponding expressions, Eqs. (\ref{Eq:JK1})-(\ref{Eq:JK4}). 
Due to the perturbative character of the latter, their sum (together with the contribution from potential exchange interaction between LMOs)  deviates from the exact value of $J$ (cf. Table \ref{Table:J}). 

\begin{table}[tb]
\begin{ruledtabular}
\caption{$J$ and its kinetic (K1-K4) and potential exchange (PE) contributions (meV).
}
\label{Table:J}
\begin{tabular}{ccccccc}
 System   & $J$   & K1   & K2      & K3      & K4     & PE\footnotemark[1] \\
\hline
 Fe-Co-Fe & $-10.4$\footnotemark[2] & 27.1 & 23.7    & $-75.1$ & $-6.2$ & $-5.1$ \\
 Cu-Cr-Cu & $-3.62$\footnotemark[2] & 0.75 & 3.14 & $-4.67$ & $-0.02$ & $-2.58$ \\
 Cu-Mo-Cu &   1.26\footnotemark[2]  & 1.14 & 4.45 & $-2.56$ & 0.00 & $-0.99$ \\
 Cu-chain & $-0.65$ & 0.49 & 0.02 & $-0.27$ & 0.00   & $-0.17$ \\
\end{tabular}
\end{ruledtabular}
\footnotetext[1]{First principles $J_\text{PE}$ is scaled down following Ref. \cite{Ku2002}.}
\footnotetext[2]{$\Delta$ was chosen to reproduce the experimental $J$.}
\end{table}

\begin{figure}[tb]
\begin{tabular}{ll}
(a) & (c) \\
\raisebox{-1.0\height}{
\includegraphics[bb=0 0 1531 711, width=3cm]{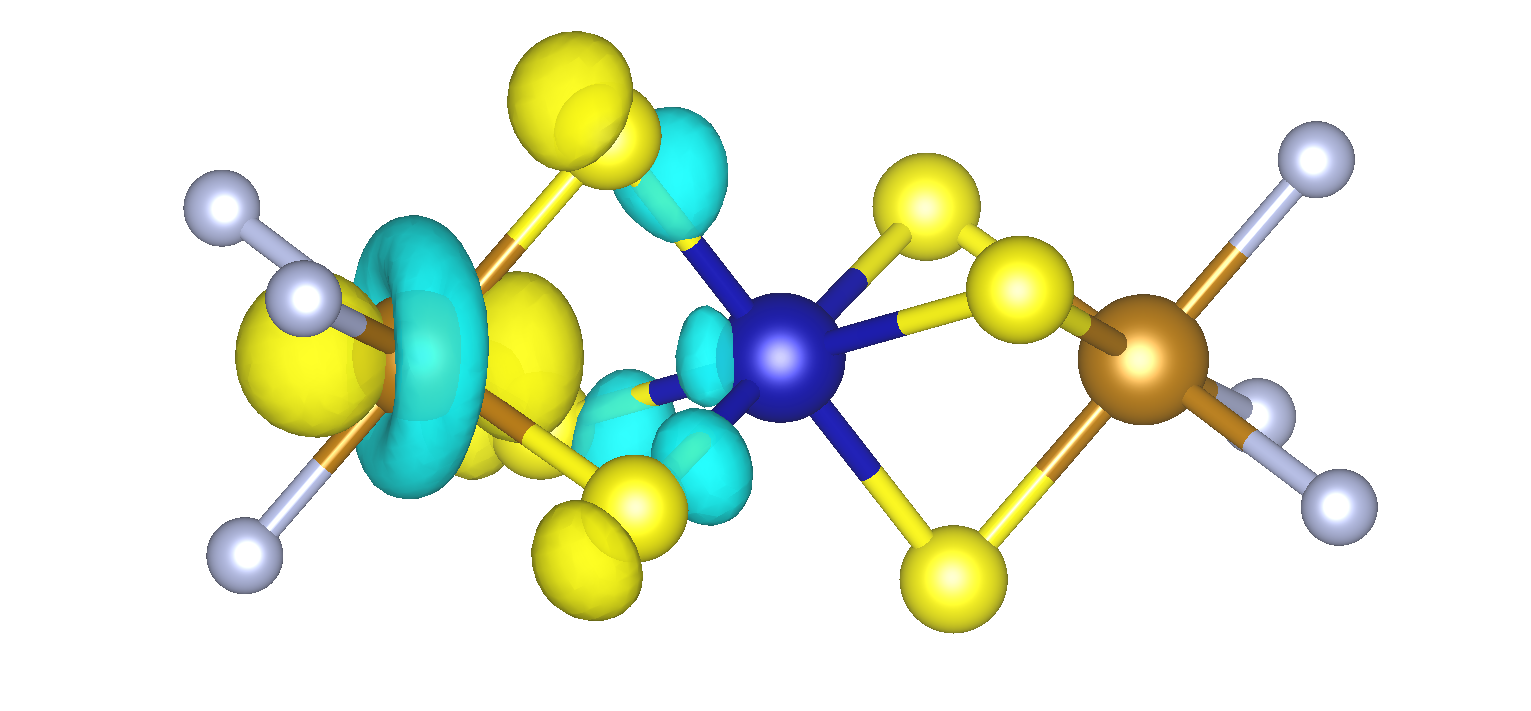}
}
&
\multirow{3}{*}{
\includegraphics[width=5cm]{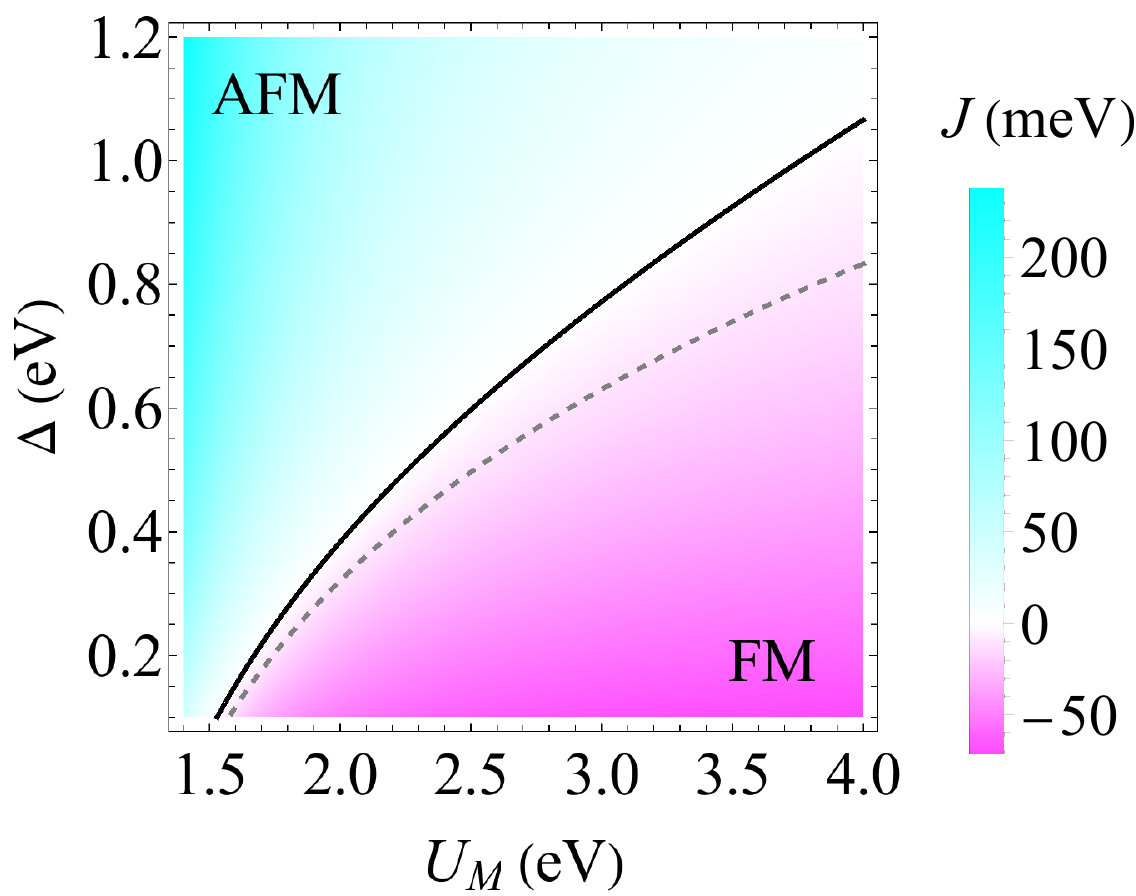}
}
\\
(b) & \\
\\
\includegraphics[bb=0 0 1531 711, width=3cm]{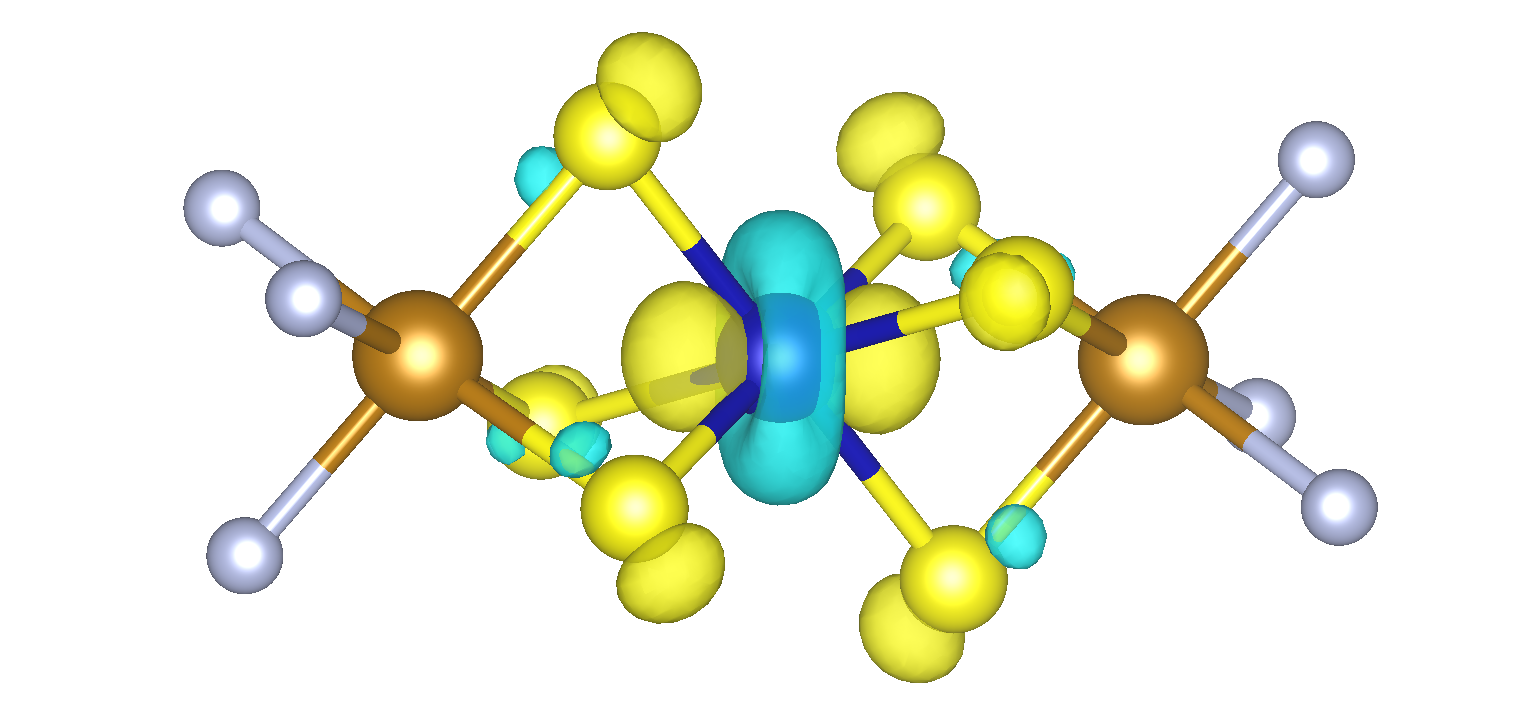} & \\
\end{tabular}
\caption{
LMO on one Fe (a) and LBO on Co (b) sites in the Fe-Co-Fe complex (only core ligand atoms are shown).
The LMO on the other Fe has the same phase as (a).
Brown, blue, yellow, and gray balls stand for Fe, Co, S and N, respectively.
(c) Exchange parameter diagram.
The solid line corresponds to $J = 0$ and the dashed line to the experimental $J$. 
}
\label{Fig:FeCoFe}
\end{figure}

\begin{figure}[bt]
\begin{tabular}{ll}
(a) & (b)
\\
\includegraphics[bb=0 0 1531 711, width=4cm]{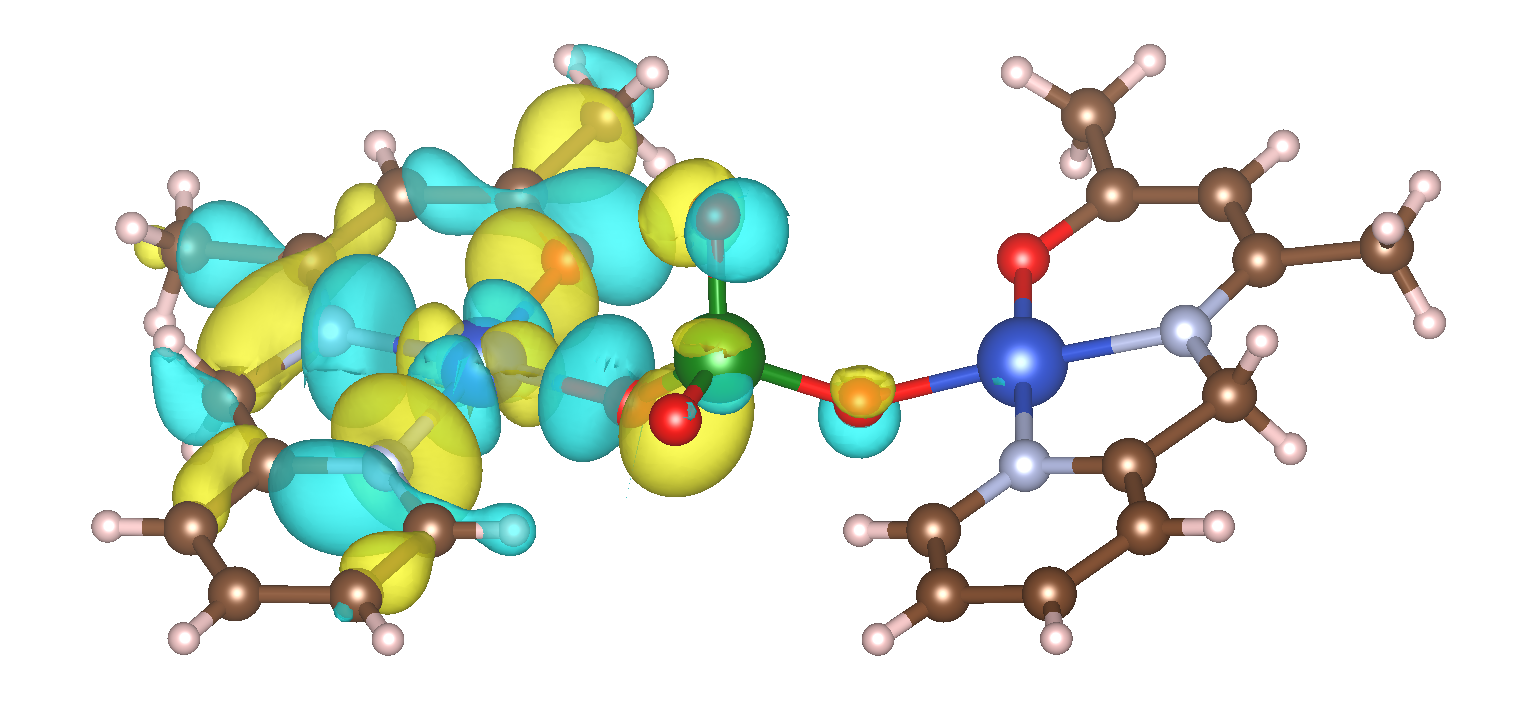}
&
\includegraphics[bb=0 0 1531 711, width=4cm]{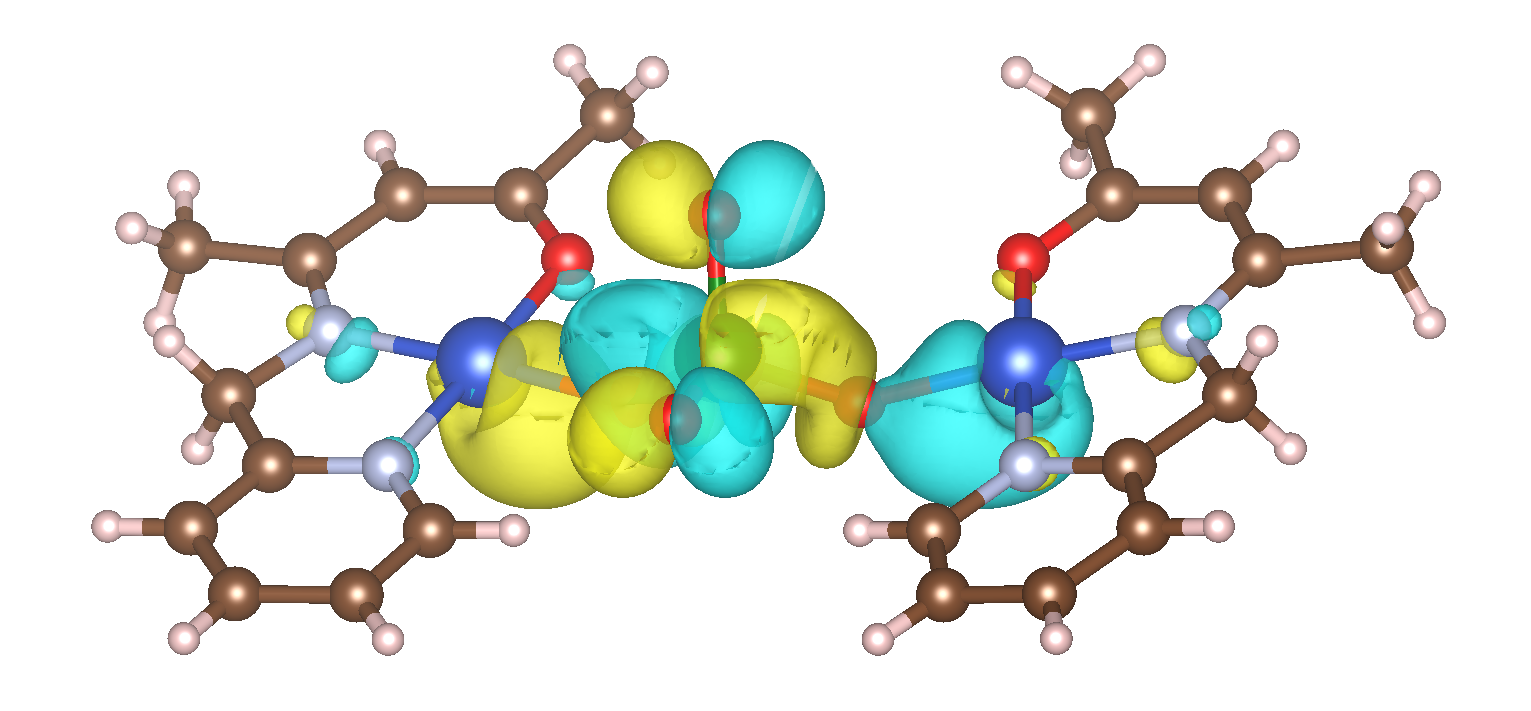}
\\
(c) & (d)
\\
\includegraphics[height=3.5cm]{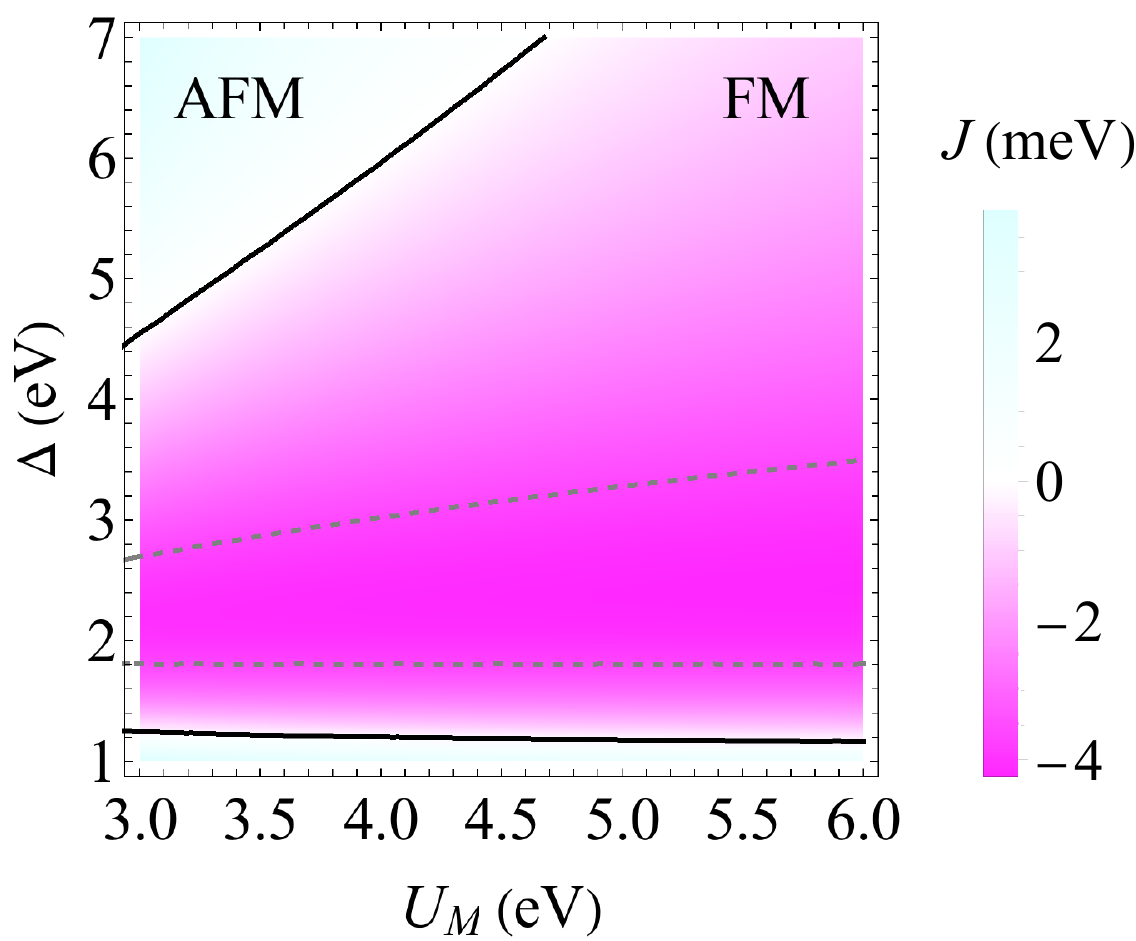}
&
\includegraphics[height=3.5cm]{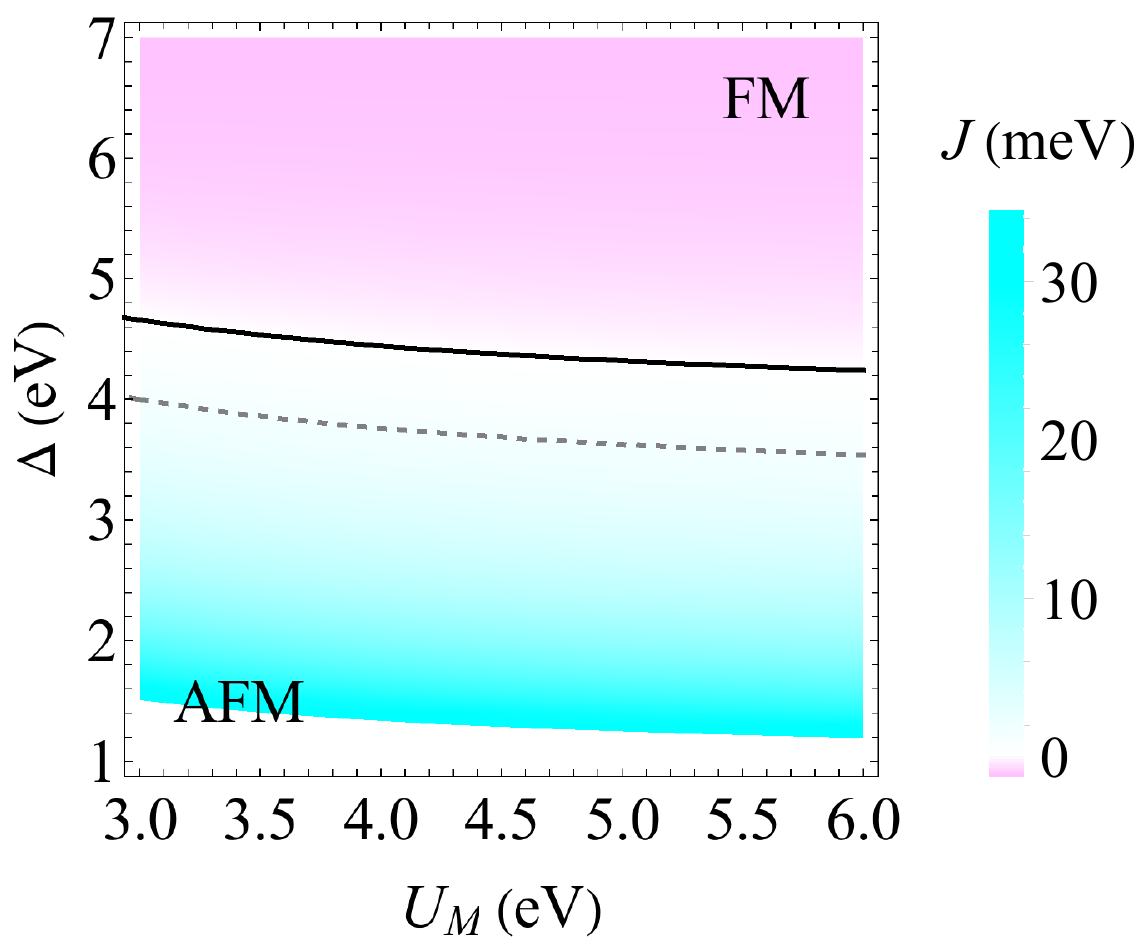}
\end{tabular}
\caption{
LMO on Cu site (a) and LBO on Cr site (b) in the Cu-Cr-Cu complex,
exchange parameter diagrams for Cu-Cr-Cu (c) and Cu-Mo-Cu (d) complexes.
The phase of the LMO at the other Cu site is opposite to (a).
The blue, green, red, light gray, dark brown, and white balls are Cu, Cr, O, N, C, and H, respectively.
The Cu-Cu axis corresponds to the $x$ axis and $z$ is the out of plane axis.
The meaning of the lines in (c) and (d) is the same as in Fig. \ref{Fig:FeCoFe}.
}
\label{Fig:CuCrCu}
\end{figure}

\begin{figure}[tb]
\centering
\begin{tabular}{lll}
(a) & (b) & (c)\\
\scalebox{-1}[-1]{\includegraphics[bb=0 0 711 711, width=0.32\columnwidth]{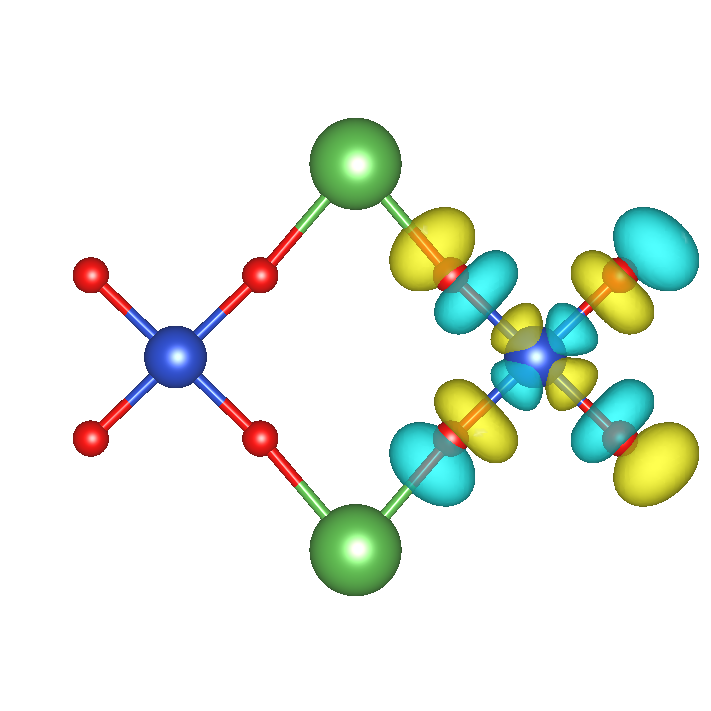}}
&
\scalebox{-1}[-1]{\includegraphics[bb=0 0 711 711, width=0.32\columnwidth]{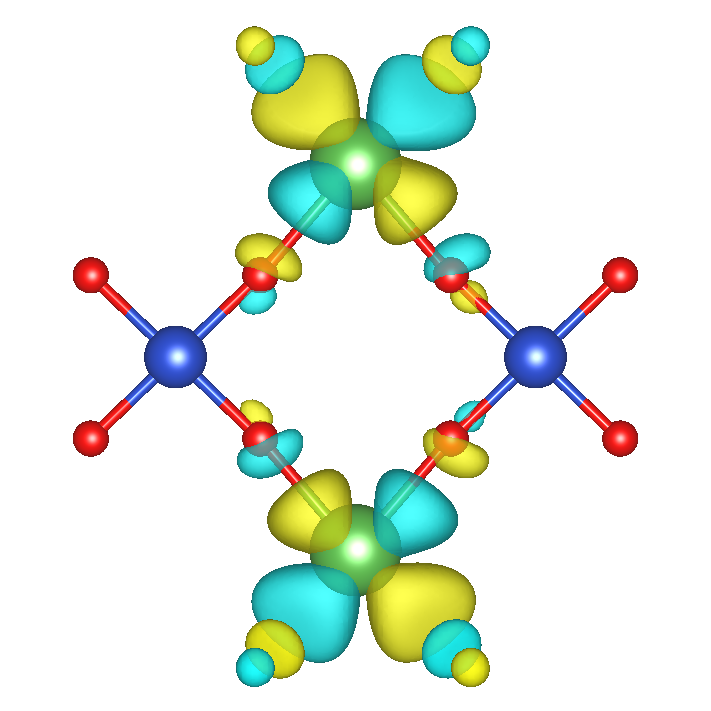}}
&
\scalebox{-1}[-1]{\includegraphics[bb=0 0 711 711, width=0.32\columnwidth]{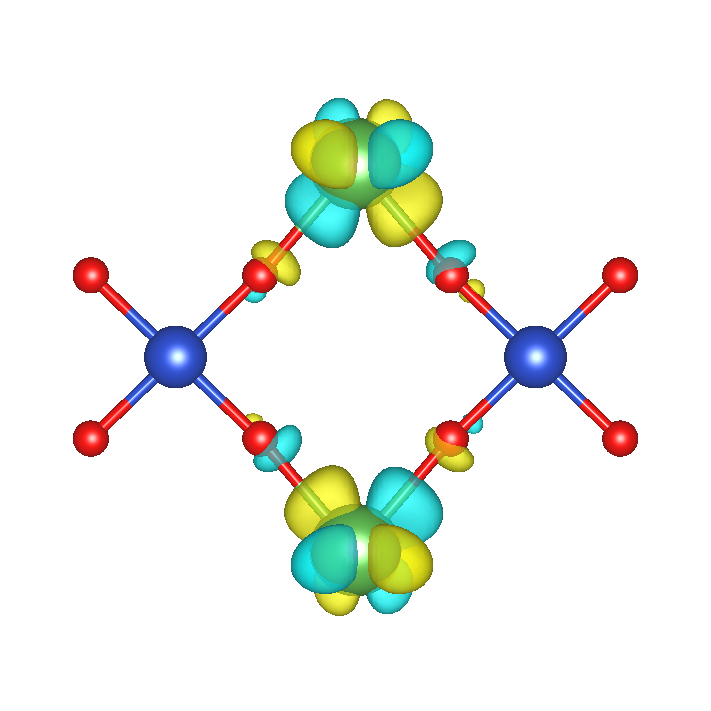}}
\\
(d) \\
\multicolumn{3}{c}{
\includegraphics[width=0.96\columnwidth]{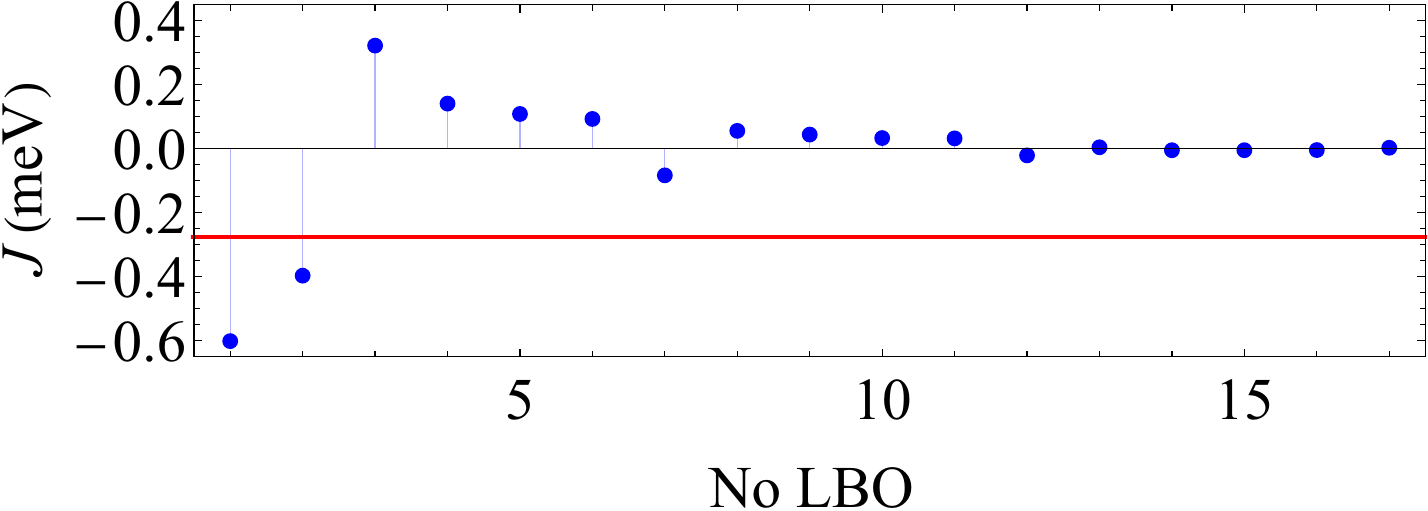}
}
\end{tabular}
\caption{
LMO on Cu (a) and two LBO on the bridging La ions giving the strongest K3 contribution (b and c). 
The phase of the LMO at the other Cu site is the same as (a).
The blue, green, and red balls correspond to Cu, La and O, respectively. 
(d) The contributions of individual LBOs to K3. The first two contributions correspond to LBO in (b) and (c), respectively.
The red line indicates the total K3 contribution from all LBOs.
}
\label{Fig:Cuchain}
\end{figure}

\subsection{Fe-Co-Fe complex}
The $3d$ orbitals of each metal ion split into $e_g$ ($e$ in $C_3$ group) and $t_{2g}$ [$a$ ($d_{z^2}$) and $e$] orbitals because of the strong octahedral-like $C_3$ ligand field. 
In both Fe and Co, the $t_{2g}$ orbitals have much lower energy than the $e_g$ orbitals and are filled by 5 and 6 electrons, respectively. 
The half-filled $a$ orbital on the Fe site is the LMO, which is consistent with the calculated spin density. 
Due to the $C_3$ symmetry, only the $a$ orbitals on Fe and Co sites are relevant to kinetic exchange interaction, while Goodenough's mechanism is ruled out. 
Below, we use the hole picture \cite{Chibotaru2003}.

The calculated LMOs and LBO [Figs. \ref{Fig:FeCoFe} (a), (b)] are strongly hybridized with the $3p$ orbitals of the sulfur atoms between the metal ions, which makes $t_{MM}$ non-negligible.  
Fig. \ref{Fig:FeCoFe}(c) shows the $J$-diagram as a function of parameters $U_M$ and $\Delta$ (the less reliable among the DFT-extracted parameters) at DFT-calculated values of other parameters (Table \ref{Table:parameter}).

The $J$-diagram shows the presence of ferromagnetism for a wide range of the parameters.
To elucidate the realistic contributions to $J$, $\Delta = 0.60$ eV was taken to reproduce its experimental value with the theoretical value of $U_M = 2.86$ eV. 
The value of $\Delta$ matches the estimation 1.05 eV from absorption spectra in solution \cite{Glaser1999}.
Table \ref{Table:J} shows that the ferromagnetic kinetic exchange (K3) is clearly dominant due to a relatively large value of $t_{MM}$ (Table \ref{Table:parameter}).
The contributions K1 and K2 are similar in magnitude because of an efficient cancellation of $t_{MD}t_{DM}/\Delta$ by $t_{MM}$ in the former. 
Thus the observed very large ferromagnetic coupling ($-10$ meV) in this complex \cite{Glaser1999} is confirmed to be entirely due to the ferromagnetic kinetic exchange mechanism.

\subsection{Cu-Cr-Cu and Cu-Mo-Cu complexes}
In tetragonal and tetrahedral environments, $3d_{x^2-y^2}$ and $3/4d_{zx}$ orbitals become LMO and LBO on Cu and Cr/Mo, respectively [Figs. \ref{Fig:CuCrCu} (a), (b)], which agrees with the calculated spin density. 
$J$-diagrams show that Cu-Cr-Cu and Cu-Mo-Cu complexes become ferro- and antiferromagnetic, respectively, for realistic $\Delta$ and $U_M$ [Figs. \ref{Fig:CuCrCu} (c), (d)].
In both complexes, due to a partial cancellation of $t_{MM}$ and $t_{MD}t_{DM}/\Delta$, the effective transfer parameter between the two LMOs, $t_{MM} - t_{MD}t_{DM}/\Delta$, is reduced and hence the K1 contribution becomes small. 
The $t_{MM}$ in Cu-Cr-Cu is larger than in Cu-Mo-Cu, and the same for the K3 contribution. 
Consequently, the former compound is ferromagnetic and the latter antiferromagnetic.

\subsection{Quasi 1D Cu chain}
The origin of ferromagnetism in La$_4$Ba$_2$Cu$_2$O$_{10}$ was debated in the past \cite{Tasaki1998, Ku2002}.
In this system, the magnetic orbitals are of $3d_{zx}$ type ($b_{2g}$) of Cu due to the tetragonal-like $D_{2h}$ ligand field [Fig. \ref{Fig:Cuchain}(a)] and the bridging orbitals are the empty orbitals of La, Ba and O, where the $c$ axis is taken as the $z$ axis and the plane of the Cu chain $zx$. 
Because of the symmetry, the Goodenough's mechanism is irrelevant: 
the irreducible representations of the other $3d$ orbitals differ from $b_{2g}$, and therefore, the electron transfers between the $3d_{zx}$ orbital and the other types of $3d$ orbitals are forbidden.

With first-principles parameters (the parameters related to the $5d_{zx}$ LBO are shown in Table \ref{Table:parameter}), we obtained $J = -0.65$ meV close to the experimental value ($-0.4$ meV \cite{Masuda1991}).
Remarkably, 
the first contribution in K2 (\ref{Eq:JK2}), which is like the K2 in Eq. (\ref{Eq:J}), is now ferromagnetic ($-0.26$) and of similar magnitude as K3 [the second (similar to K1) and the third (resembling K3) contributions are 0.18 and 0.09, respectively]. 
The first term of K2 $<$ 0 became possible due to numerous loop terms involving two different LBOs [the third term in Eq. (\ref{Eq:JK2})] which can be negative when $t_{MD}=t_{DM}$ for one LBO and $t_{MD}=-t_{DM}$ for the other. 
For the same reason both ferro- and antiferromagnetic contributions for different LBOs are present in Eq. (\ref{Eq:JK3}) reducing the total K3 contribution [Fig. \ref{Fig:Cuchain}(d)]. 
Among the latter, the contributions via the $5d_{zx}$ and the $4f_{z(x^2-y^2)}$ of in-plane La ions [Figs. \ref{Fig:Cuchain}(b),(c)] are dominant.
Thus two kinetic ferromagnetic exchange mechanisms, K2 and K3, make together a dominant contribution rendering the resulting exchange interaction ferromagnetic.

The potential exchange interaction between the Anderson's magnetic orbitals was attributed to the origin of ferromagnetism based on the Anderson's original approach \cite{Ku2002} (see for further discussion Appendix \ref{A:JPE}), while the present analysis shows that the ferromagnetic kinetic contributions which is missing in the Anderson's model (\ref{Eq:HAMO}) is more important.

\subsection{Fingerprint of ferromagnetic kinetic mechanism}
There is further evidence of the dominant contribution of the ferromagnetic kinetic exchange mechanism in the Fe complex and the Cu chain.
As described in Sec. \ref{Sec:switch}, this mechanism is quenched when replacing the bridging metal ion with a magnetic one. 
Such behavior was observed in a series of trinuclear isostructural Fe complexes with various electronic populations of the central metal ion \cite{Glaser1999}
and between Cu ions in La$_4$Ba$_2$Cu$_2$O$_{10}$ under the substitution of the diamagnetic La$^{3+}$ by the paramagnetic Nd$^{3+}$ \cite{Paukov1991, Golosovsky1993}.

\section{Conclusions}
We have investigated the ferromagnetic kinetic exchange interaction between localized spins. This mechanism shows up at a higher level of treatment compared to Anderson's theory, through the separation and explicit consideration of relevant diamagnetic orbitals bridging the magnetic ones. The crucial point is that despite a stronger localization compared to AMOs, the LMOs and LBOs arising in the present treatment are by far not atomic like. This opens two paths for delocalization of magnetic electrons, via the LBOs and through-space. When the latter is sufficiently strong, the interference between the two kinetic paths can result in a ferromagnetic contribution which overcomes the conventional antiferromagnetic superexchange. The conditions for achieving strong ferromagnetism via this mechanism have been elucidated. In particular, it is favored by the reduced orbital gap between magnetic and bridging orbitals, pointing to materials with strong metal-ligand covalency. 

We have investigated the relevance of the ferromagnetic kinetic exchange mechanism in several compounds by first-principles calculations. 
It was found that this exchange contribution is of comparable magnitude with the antiferromagnetic kinetic exchange. The calculations show that in the Fe-Co-Fe complex the observed very large ferromagnetic coupling is entirely due to a strong ferromagnetic kinetic contribution.
We also discovered a fingerprint of the ferromagnetic kinetic mechanism: a switching of the ferromagnetism to antiferromagnetism by substituting the diamagnetic bridging site with magnetic one. The phenomenology is observed in two series of systems, supporting the importance of the mechanism in magnetic materials.
The obtained results call for the reconsideration of the origin of ferromagnetism and weak antiferromagnetism in insulating magnetic materials and complexes.

\begin{acknowledgments}
Z.H. and D.L. were supported by the China Scholarship Council.
A.M. acknowledges funding provided by the Magnus Ehrnrooth Foundation.
V.V. was postdoctoral fellow of the Research Foundation - Flanders (FWO).
N.I. was partly supported the GOA program of KU Leuven and the scientific research grant R-143-000-A80-114 of the National University of Singapore.
The computational resources were provided by the VSC (Flemish Supercomputer Center). 
\end{acknowledgments}

\appendix

\section{A derivation of $J$}
We show an alternative derivation of the exchange parameter (\ref{Eq:J}).
We introduce ferromagnetic (upper) and antiferromagnetic (lower) configurations,
\begin{eqnarray}
|\Phi^\pm_{ij}\rangle = \frac{1}{\sqrt{2}}(\hat{a}^\dagger_{i\uparrow} \hat{a}^\dagger_{j\downarrow} \pm \hat{a}^\dagger_{i\downarrow} \hat{a}^\dagger_{j\uparrow})|0\rangle,
\label{Eq:Phi}
\end{eqnarray}
where $i,j = 1,2,d$, and $|0\rangle$ is the vacuum state.
The lowest energy configurations $|\Phi^\pm_{12}\rangle$ slightly hybridize with $|\Phi^\pm_\text{ex} \rangle = \frac{1}{\sqrt{2}} (\tau |\Phi_{1d}^\pm\rangle + |\Phi_{d2}^\pm\rangle)$, due to the electron transfer interaction between the magnetic and diamagnetic sites, where $\tau = t_{DM}/t_{MD}$. 
Taking the electron transfer interaction as the perturbation and the rest in Eq. (\ref{Eq:H}) as the unperturbed Hamiltonian, the ground states are calculated within second order perturbation theory as, 
\begin{eqnarray}
|\Psi^\pm\rangle = |\Phi^\pm_{12}\rangle - \frac{\sqrt{2}t_{MD}}{\Delta \mp \tau t_{MM}} |\Phi^\pm_\text{ex} \rangle + \cdots.
\label{Eq:Psi}
\end{eqnarray}
where $\Delta \mp \tau t_{MM}$ is the energy of $|\Phi^\pm_\text{ex} \rangle$, and the terms which are not directly relevant to the K3 term (\ref{Eq:J}) are not explicitly written. 
The energies with respect to $|\Phi^\pm\rangle$ (\ref{Eq:Psi}) are calculated as 
\begin{eqnarray}
 E^\pm &=& E^\pm_0 -\frac{2t_{MD}^2}{\Delta \mp \tau t_{MM}}
\\
 &\approx & E^\pm_0 - \frac{2t_{MD}^2}{\Delta} \pm \frac{2t_{MD}t_{DM}t_{MM}}{\Delta^2}.
\label{Eq:Epm}
\end{eqnarray}
The last term of Eq. (\ref{Eq:Epm}) corresponds to the K3 contribution, and $E_0^\pm$ contains all the other contributions. 

The ferromagnetic contribution arises by the spin dependent covalency between paramagnetic centers.
$|\Phi^\pm_\text{ex} \rangle$ is transformed as 
\begin{eqnarray}
 |\Phi^\pm_\text{ex} \rangle
 &=& 
  \frac{1}{2}
  \left[
  (\tau \hat{a}^\dagger_{1\uparrow} \mp \hat{a}^\dagger_{2\uparrow}) \hat{a}^\dagger_{d\downarrow}
 \pm
  (\tau \hat{a}^\dagger_{1\downarrow} \mp \hat{a}^\dagger_{2\downarrow}) \hat{a}_{d\uparrow}^\dagger 
  \right]
 |0\rangle.
\quad
 \label{Eq:Phiex}
\end{eqnarray}
Note that the molecular orbital states, $(\tau \hat{a}^\dagger_{1\sigma} \mp \hat{a}^\dagger_{2\sigma})/\sqrt{2}$, depend on the total spin (triplet or singlet), and consequently, their orbital energy levels ($\mp \tau t_{MM}$) too.

A mathematical discussion on the relation between the symmetry of the wave function and the ferromagnetic ground state is found in Ref. \cite{Tasaki1998Hubbard}

\section{Anderson's model}
\label{A:Anderson}
The tight-binding model (\ref{Eq:HAMO}) is derived as follows \cite{Anderson1959}.
(1) Calculation of the molecular orbitals for the high-spin state within restricted open-shell Hartree-Fock calculations;
(2) transformation of the magnetic molecular orbitals (a symmetric and an antisymmetric half-filled molecular orbitals in the present case) into the localized orbitals (AMOs). 
The other molecular orbitals are ignored.
(3) Transformation of the Hamiltonian in the space of the AMOs. 

From the two LMOs and one LBO in the basic three-site model, two symmetric ($S$, $S'$) and one antisymmetric ($A$) molecular orbitals are constructed as 
\begin{eqnarray}
 \hat{c}_{S\sigma}^\dagger &=& \frac{\cos\theta}{\sqrt{2}} \left(\hat{a}_{1\sigma}^\dagger + \tau \hat{a}_{2\sigma}^\dagger\right) + \sin \theta \hat{a}_{d\sigma}^\dagger,
\label{Eq:MOS1}
\\
 \hat{c}_{S'\sigma}^\dagger &=& -\frac{\sin\theta}{\sqrt{2}} \left(\hat{a}_{1\sigma}^\dagger + \tau \hat{a}_{2\sigma}^\dagger\right) + \cos \theta \hat{a}_{d\sigma}^\dagger,
\label{Eq:MOS2}
\\
 \hat{c}_{A\sigma}^\dagger &=& \frac{1}{\sqrt{2}}\left(\hat{a}_{1\sigma}^\dagger - \tau \hat{a}_{2\sigma}^\dagger\right),
\label{Eq:MOA}
\end{eqnarray}
where $\tau = t_{DM}/t_{MD}$,
We assume $\cos\theta > |\sin\theta|$ and the orbital energy for the state $S$ is lower than that for the state $S'$.
The angle $\theta$ is determined so that the high-spin state energy becomes the minimum within the restricted open-shell Hartree-Fock method.

Anderson's magnetic orbitals are defined by using partially filled orbitals:
\begin{eqnarray}
 \hat{A}_{1\sigma}^\dagger &=& \frac{1}{\sqrt{2}}\left(\hat{c}_{S\sigma}^\dagger + \hat{c}_{A\sigma}^\dagger \right),
\label{Eq:A1}
\\
 \hat{A}_{2\sigma}^\dagger &=& \frac{1}{\sqrt{2}}\left(\hat{c}_{S\sigma}^\dagger - \hat{c}_{A\sigma}^\dagger \right).
\label{Eq:A2}
\end{eqnarray}
In Anderson's approach, the other orbital $\hat{c}_{S'\sigma}$ is omitted. 
Thus, within this approximation, the atomic orbitals are expressed as 
\begin{eqnarray}
 \hat{a}_{1\sigma}^\dagger
 &=& 
 \frac{\cos\theta+1}{2} \hat{A}_{1\sigma}^\dagger + \frac{\cos\theta-1}{2} \hat{A}_{2\sigma}^\dagger,
\label{Eq:a1AMO}
\\
 \hat{a}_{2\sigma}^\dagger
 &=&
 \tau \frac{\cos\theta-1}{2} \hat{A}_{1\sigma}^\dagger + \tau \frac{\cos\theta+1}{2} \hat{A}_{2\sigma}^\dagger,
\label{Eq:a2AMO}
\\
 \hat{a}_{d\sigma}^\dagger 
 &=&
 \frac{\sin\theta}{\sqrt{2}} \left(\hat{A}_{1\sigma}^\dagger + \hat{A}_{2\sigma}^\dagger\right).
\label{Eq:adAMO}
\end{eqnarray}

Substituting Eqs. (\ref{Eq:a1AMO})-(\ref{Eq:adAMO}) into the single electron part $\hat{H}_1$ of Eq. (\ref{Eq:H}), 
\begin{eqnarray}
 \hat{H}_1
 &=& 
 E_\text{HF}
 + 
 \sum_{\sigma} 
  \tau b
  \left(\hat{A}_{1\sigma}^\dagger \hat{A}_{2\sigma} + \hat{A}_{2\sigma}^\dagger \hat{A}_{1\sigma}\right),
 \label{Eq:H1AMO}
\end{eqnarray}
where $E_\text{HF}$ is defined by 
\begin{eqnarray}
 E_\text{HF} &=& \frac{\Delta - \tau t_{MM}}{2} (1 - \cos 2\theta) + \sqrt{2} t_{MD} \sin 2\theta
\nonumber\\
 &=& \frac{\Delta - \tau t_{MM}}{2} - R \cos (2\theta + \alpha),
 \label{Eq:EHF0}
\end{eqnarray}
$R$ and $\alpha$ are 
\begin{eqnarray}
  R &=& \sqrt{\left(\frac{\Delta - \tau t_{MM}}{2}\right)^2 + \left(\sqrt{2}t_{MD}\right)^2},
\label{Eq:R}
\\
 \cos\alpha &=& \frac{\Delta - \tau t_{MM}}{2R},
\quad
 \sin\alpha = \frac{\sqrt{2} t_{MD}}{R},
\label{Eq:alpha}
\end{eqnarray}
respectively, 
and $b$ is 
\begin{eqnarray}
 b = t_{MM} + \tau \frac{E_\text{HF}}{2}. 
\label{Eq:bAMO}
\end{eqnarray}
The angle $\theta$ in $E_\text{HF}$ is fixed below. 
Since there are two electrons in total, $\sum_i \sum_\sigma \hat{N}_{i\sigma}$ was replaced by 2. 

Similarly, substituting 
\begin{eqnarray}
 \hat{n}_{1(2) \sigma} &=& 
 \frac{1}{4}\left[(\cos\theta+1)^2 \hat{N}_{1(2)\sigma} + (\cos\theta-1)^2 \hat{N}_{2(1)\sigma}
 \right.
\nonumber\\
 &&-
 \left.
  \sin^2\theta \left(\hat{A}_{1\sigma}^\dagger \hat{A}_{2\sigma} + \hat{A}_{2\sigma}^\dagger \hat{A}_{1\sigma} \right)
 \right],
 \label{Eq:n12AMO}
\\
 \hat{n}_{d\sigma} &=& 
 \frac{1}{2}
  \sin^2\theta 
 \left[
  \hat{N}_{1\sigma} + \hat{N}_{2\sigma} 
  +
  \left( \hat{A}_{1\sigma}^\dagger \hat{A}_{2\sigma} + \hat{A}_{2\sigma}^\dagger \hat{A}_{1\sigma} \right)
 \right],
\nonumber\\
 \label{Eq:ndAMO}
\end{eqnarray}
into the on-site Coulomb terms in Eq. (\ref{Eq:H}), 
\begin{eqnarray}
 \hat{H}_\text{Coul}
&=& 
 \sum_{i=1,2} U \hat{N}_{i\uparrow} \hat{N}_{i\downarrow}
 +
 \sum_{\sigma}
 b' \left(\hat{N}_{1,-\sigma} + \hat{N}_{2,-\sigma} \right) 
\nonumber\\
 && \times
  \left(\hat{A}_{1\sigma}^\dagger \hat{A}_{2\sigma} + \hat{A}_{2\sigma}^\dagger \hat{A}_{1\sigma} \right)
 + U' \left[ \hat{N}_{1\uparrow} \hat{N}_{2\downarrow} + \hat{N}_{2\uparrow} \hat{N}_{1\downarrow} 
 \right.
\nonumber\\
 && + 
 \left.
  \left(\hat{A}_{1\uparrow}^\dagger \hat{A}_{2\uparrow} + \hat{A}_{2\uparrow}^\dagger \hat{A}_{1\uparrow} \right)
  \left(\hat{A}_{1\downarrow}^\dagger \hat{A}_{2\downarrow} + \hat{A}_{2\downarrow}^\dagger \hat{A}_{1\downarrow} \right)
 \right],
\nonumber\\
 \label{Eq:HCoulAMO}
\end{eqnarray}
where $U$, $U'$, and $b'$ are, respectively, defined by 
\begin{eqnarray}
 U &=& \frac{U_M}{8}
  \left(\cos^4\theta + 6 \cos^2\theta + 1\right) 
 + 
 \frac{U_D}{4} \sin^4\theta,
\label{Eq:UAMO}
\\
 U' &=& \left( \frac{U_M}{8} + \frac{U_D}{4} \right) \sin^4 \theta,
\label{Eq:U'}
\\
 b' &=& -\frac{U_M}{8} (\cos^2\theta + 1) \sin^2\theta + \frac{U_D}{4} \sin^4\theta.
\label{Eq:b'}
\end{eqnarray}
Due to the assumption on $\theta$, $\cos \theta > |\sin \theta|$, $U > U'$.
The Coulomb term contains on-site and intersite Coulomb interactions, pair electron transfer, and Coulomb assisted electron transfer interactions.
The model Hamiltonian including all terms in Eqs. (\ref{Eq:H1AMO}) and (\ref{Eq:HCoulAMO}) is an extended version of Anderson's tight-binding model (\ref{Eq:HAMOgen}). 
The model omitting the terms except for $E_\text{HF}$, $b$ and $U$ is the original Anderson model.

The expectation value of the Hamiltonian for $|\psi_\text{HF}\rangle = \hat{c}_{S\uparrow}^\dagger \hat{c}_{A\uparrow}^\dagger |0\rangle = \hat{A}_{1\uparrow}^\dagger \hat{A}_{2\uparrow}^\dagger|0\rangle$ corresponds to $E_\text{HF}$ (\ref{Eq:EHF0}). 
The energy becomes minimum,
\begin{eqnarray}
 E_\text{HF} &=& \frac{\Delta - \tau t_{MM}}{2} - R,
\label{Eq:EHF}
\end{eqnarray}
when 
\begin{eqnarray}
 2\theta + \alpha = 2\pi n, \quad n \in \mathbb{Z},
\label{Eq:theta}
\end{eqnarray}
or 
\begin{eqnarray}
 \cos 2\theta = \cos \alpha, \quad \sin 2\theta = -\sin \alpha. 
\end{eqnarray}
Choosing such $\theta$, we obtain the extended Anderson model. 

From the model, we obtain triply degenerate ferromagnetic (spin triplet) states and three nondegenerate low-spin (singlet) states. 
The energy of the level for the high-spin states corresponds to $E_\text{HF}$ (\ref{Eq:EHF}).
On the other hand, the ground energy level for the antiferromagnetic states is 
\begin{eqnarray}
 E_\text{AF} &=& E_\text{HF} + \frac{U+3U'}{2} - \sqrt{\left(\frac{U-U'}{2}\right)^2 + 4(b+b')^2}.
\nonumber\\
 \label{Eq:ELSAnderson}
\end{eqnarray}
Thus, the Heisenberg exchange parameter (\ref{Eq:HHeisenberg}) is 
\begin{eqnarray}
 J &=& -\frac{U+3U'}{2} + \sqrt{\left(\frac{U-U'}{2}\right)^2 + 4(b+b')^2}.
\label{Eq:JAMO}
\end{eqnarray}

\section{DFT calculations}
\label{A:DFT}
\subsection{Fe-Co-Fe}
In the DFT calculation, the plane-wave kinetic energy cutoff was set to 100 Ry with the density cut-off of 400 Ry, and the $\Gamma$ point was used to perform the Brillouin zone integration for both self-consistent and non-self-consistent calculations.
The convergence criterion of the total energy was set to be 10$^{-10}$ Hartree.

In the constrained random phase approximation (cRPA) calculation, the energy cutoff of the polarization function was set 10 Ry.
For the convergence of the calculations of Coulomb and exchange parameters, the polarization effects from 1200 bands (338 occupied bands, and 862 unoccupied bands) were included.
The electronic energy bands of the complexes are nearly flat, and we choose three relevant bands (332nd, 338th, 339th) to generate maximally localized Wannier functions. 

The $t_{MM/MD}$ do not depend much on the choice of the functional and conditions of the calculations: The local density approximation values are 0.269 and 0.189 eV with loose conditions (energy cut-off is 50 Ry, density cut-off is 200 Ry, and threshold of the total energy change is 10$^{-6}$ Hartree).

\subsection{Cu-Cr-Cu and Cu-Mo-Cu}
Almost all the conditions for the band and cRPA calculations of the Cu complexes are the same as those for Fe-Co-Fe.
The Wannier orbitals were generated by using 111st, 112nd, and 126th bands for Cu-Cr-Cu and the 111st, 112nd, and 131st for Cu-Mo-Cu.
The total number of bands for the screened Coulomb and exchange parameters was 650 (111 occupied bands, and 539 unoccupied bands) in both cases.

\subsection{Ba$_4$La$_2$Cu$_2$O$_{10}$}
In the band calculation, the plane-wave kinetic energy cutoff was set to 100 Ry with the density cut-off of 400 Ry, and $7\times7\times8$ Monkhorst-Pack meshes was used to perform Brillouin zone integration in order to ensure the convergence of the results.
The convergence of the total energy was set to be better than 10$^{-8}$ Hartree.

\begin{figure*}[tb]
\begin{tabular}{lll}
(a) &~& (b) \\
\includegraphics[width=0.95\columnwidth]{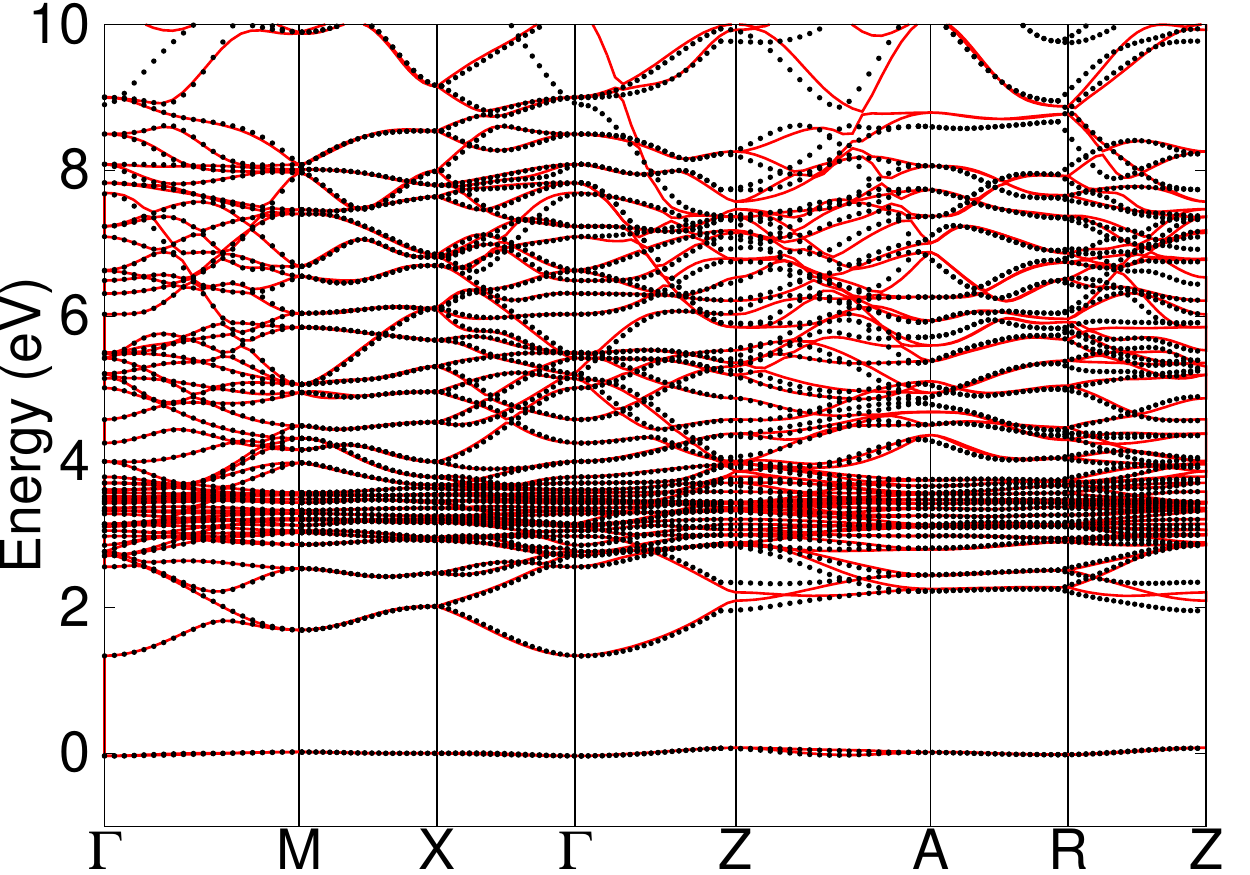}
& &
\includegraphics[width=0.95\columnwidth]{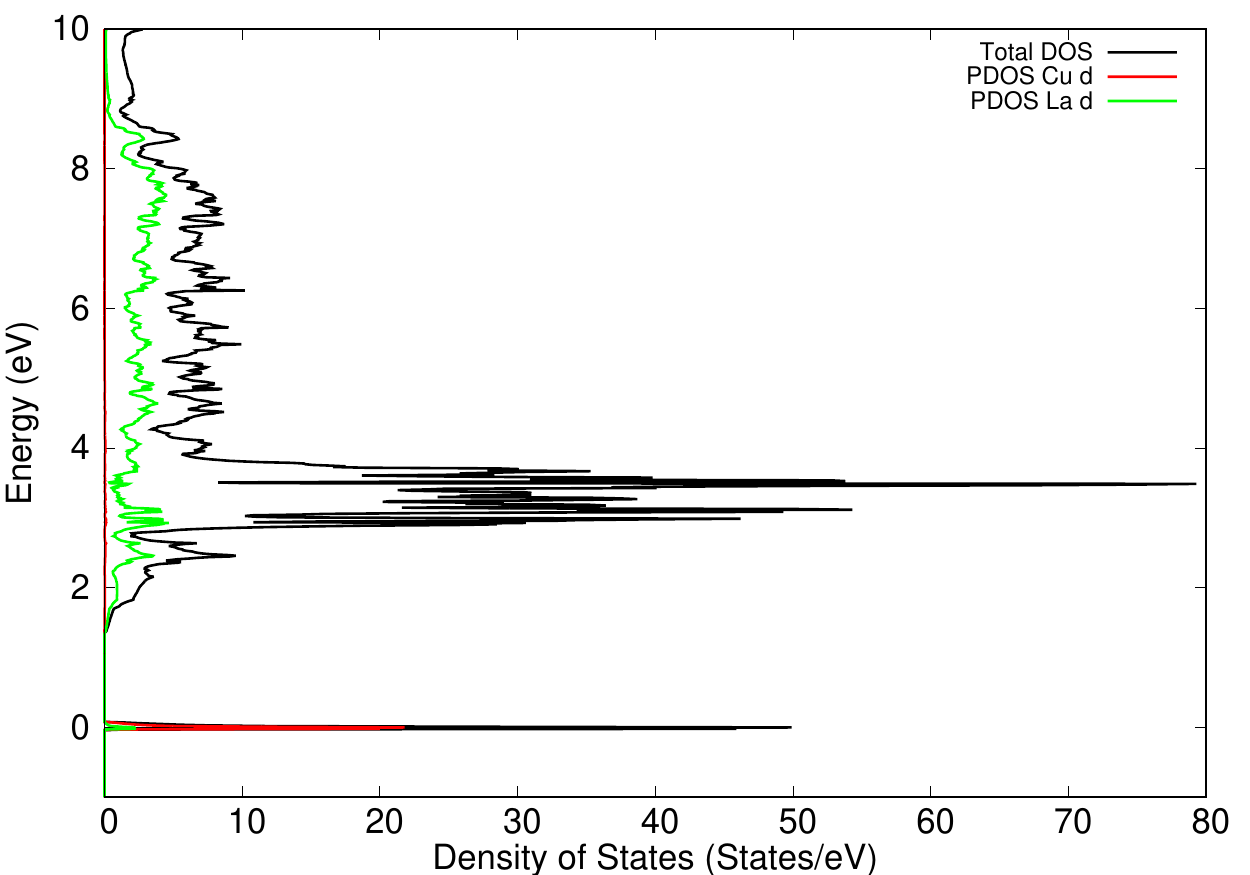}
\end{tabular}
\caption{
(a) Electronic energy band (eV) and (b) DOS and PDOS of La$_4$Ba$_2$Cu$_2$O$_{10}$.
The Fermi energy is chosen as the origin of the energy.
The black points and the red lines are the DFT values and fitting using the tight-binding Hamiltonian in the Wannier orbitals basis, respectively.
}
\label{Fig:La4Ba2Cu2O10}
\end{figure*}

Contrary to the cases of complexes, the bands originating from the bridging sites of Ba$_4$La$_2$Cu$_2$O$_{10}$ are highly complex (Fig. \ref{Fig:La4Ba2Cu2O10}).
The projected density of states (PDOS) show that the Fermi level (0 eV) is at the $d$ band as expected.
The empty $4f$ bands of La appear about 3-4 eV above the Fermi level, and the $5d$ orbital largely spread to the DOS between $-0.28$ and 10.21 eV due to the large spatial delocalization.
Therefore, 66 bands were included to generate maximally localized Wannier orbitals (see Appendix \ref{A:DFT}).
To construct 66 Wannier orbitals, a $4\times4\times4$ Monkhorst-Pack meshes of the Brillouin zone was used.
In the cRPA calculations, the energy cutoff of the polarization function was set to 10 Ry.
For the convergency of the calculations of the screened Coulomb and exchange parameters 700 bands (80 occupied, 2 partially-occupied, and 618 unoccupied bands) were included.

\section{Potential exchange interaction}
\label{A:JPE}
In the previous works based on Anderson's orbitals \cite{Ku2002, Mazurenko2007}, the potential exchange interaction was considered to originate from the Hund's rule coupling between different orbitals on the bridging site.
This contribution appears in the present approach partly as the potential exchange interaction (\ref{Eq:JPE}) and partly as Goodenough's contribution in the last term of K2 (\ref{Eq:JK2}).
Since the Hund's coupling shifts the activation energy in the denominator of K2 (\ref{Eq:JK2}) as in usual Goodenough's mechanism, the latter contribution is estimated as 
\begin{eqnarray}
  -\eta \times (\text{the third term of }J_{\text{K}2}),
\label{Eq:JPE2}
\end{eqnarray}
where $\eta$ is defined by 
\begin{eqnarray}
 \eta = \frac{J_{dd'}}{\Delta_d+\Delta_{d'}-V_{MM}+V_{dd'}}.
\end{eqnarray}
According to our calculations of La$_4$Ba$_2$Cu$_2$O$_{10}$, $\Delta \approx 5$ eV, $V_{MM} \approx 0.5$ eV, $V_{dd'} \approx 1.5$ eV and $J_{dd'} \approx 0.2$ eV for $d = 5d_{zx}$ and $d' = 4f_{z(x^2-y^2)}$ [Figs. \ref{Fig:Cuchain} (b) and (c)], and thus, $\eta \approx 0.02$.
Therefore, this Goodenough-type contribution (\ref{Eq:JPE2}) is by 2 order of magnitude weaker than the contribution (\ref{Eq:JPE}).


%

\end{document}